\documentclass{ws-book9x6}
\usepackage{ws-book-har}           

\title{My Book Title}              

\begin{document}

\author{Marina Samoylova}
\author{Nicola Piovella}
\author{Michael Holynski}
\author{Philippe Wilhelm Courteille}
\author{Romain Bachelard}

\titlepages                        



\tableofcontents


\chapter[One-dimensional photonic band gaps in optical lattices]{One-dimensional photonic band gaps in optical lattices\label{ch1}}

\section{Introduction\label{intro}}

While the scattering of light by a single point-like particle (i.e., one much smaller than the wavelength of the light) is a well-known process, the scattering by a dense assembly of particles is a rich and still open field. The {\it cooperation} of particles in scattering of coherent light was first described by Dicke~\cite{Dicke1954}, and subsequently termed Dicke superradiance: provided the density is significant, each particle reemits a wave that is in phase with those emitted by its neighbours, resulting in a constructive interference phenomenon.

The advent of laser cooling has led to the development of cold atomic physics and boosted the interest in coherent, superradiant phenomena. Low temperatures provide almost motionless particles thus avoiding Doppler broadening, an important contribution to decoherence, and facilitating cooperation between close particles. At even lower temperature, Bose-Einstein condensates correspond to a state in which all the particles phase match, with an overall very low momentum spread.

For these cold systems, the cooperativity corresponds to the capacity of the atoms to affect the radiation, i.e., for a high cooperativity each atom receives a significant quantity of radiation from the other atoms, in addition to the external laser source. High optical densities can, for example, give rise to inhomogeneities in the radiation pressure force acting on the atoms,  resulting in distortion of the cloud~\cite{Walker1990,Sesko1991}. The phenomenon of collectivity should be considered an asset as it is a coherent effect that can be controlled: for instance, inhomogeneities in the radiation pressure field have been used to define a new compression scheme~\cite{Khaykovich1999}.

Among the collective effects of light scattering, {\it single photon} scattering by an ensemble of atoms is particularly fascinating. Superradiance of single photons is possible due to the quantum state of the system where each atom has a probability to absorb the unique photon present in the system, and the {\it synchronized} radiation of the atoms is at the origin of the superradiance. For example, in the case of a cloud of atoms prepared in the ``timed Dicke'' state, cooperativity makes the atoms emit a superradiant light pulse in a narrow forward cone~\cite{Scully2009}. This single-photon cooperative process also gives rise to a collective Lamb shift~\cite{Rohlsberger2010,Scully2010,Keaveney2012}, as well as to other collective frequency shifts~\cite{Friedberg1973}.

Cooperative single photon scattering has been studied experimentally with ultra-cold atomic clouds~\cite{Bux2010,Bienaime2010,Bender2010}, where cooperativity manifests in a collapse or an increase of the radiation pressure force acting on the center of mass~\cite{Courteille2010}. This effect is particularly strong when resonance conditions known as Mie resonances are satisfied~\cite{Bachelard2012}. In general, both the scattered radiation intensity and the radiation pressure force contain signatures of collective scattering and each can serve as a powerful experimental tool for investigating the role of cooperativity in the atom-radiation interaction~\cite{Bachelard2012a,Chabe2013}.

The greatest potential of using ultra-cold atomic clouds to scatter light may reside in {\it ordered} configurations. A typical tool for analyzing periodic structures is Bragg scattering, the diffraction pattern resulting from the interference of the radiation emitted by a periodic array of scatterers excited by an incident light field. This has been used to study various periodic systems, such as crystals, molecules or even artificial photonic bandgap materials. Periodic lattice geometries can also be realized with ultra-cold atomic ensembles by confining the atoms within an optical standing wave. If the array is large enough the interference pattern is constructive only in the precisely defined directions that satisfy the Bragg condition. In this sense Bragg scattering is a particular case of cooperative scattering~\cite{Friedberg1973,Scully2006,Courteille2010} from an ordered structure in the limit of small optical densities, where every photon is scattered once at most. In this limit, the Bragg radiation pattern actually corresponds to the structure factor of the atomic cloud.

\begin{figure}[h]
    \centerline{\psfig{file=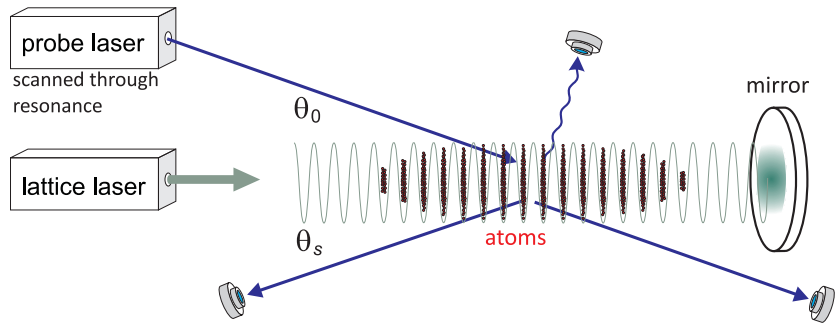,width=11cm}}
    \caption{Experimental scheme to study light scattering on one-dimensional optical lattices. A far
        red detuned retro-reflected {\it lattice laser} creates a standing wave which acts as a
        one-dimensional periodic trap for the atoms. A {\it probe laser} is used to probe the cooperative scattering
        properties of the lattice in the vicinity of the atomic resonance.}
    \label{fig:ExpScheme}
\end{figure}

Bragg scattering experiments using optical lattices have been performed (see Fig.~\ref{fig:ExpScheme}) in the thin grating regime, where the optical density of the cloud is so low that multiple light scattering events between the stacked atomic layers are rare. In the thick grating regime, characterized by multiple reflections of the incident light, application of specific light frequencies or irradiation angles is expected to give rise to {\it photonic band gaps} (PBG) also known as {\it stop bands} or {\it forbidden bands}. Clearly the collective scattering picture conserves its validity in the optically dense regime of multiple scattering and cooperative effects are responsible for the formation of forbidden photonic bands~\cite{Samoylova2013}.

Photonic band gaps for atomic clouds in optical lattices have been predicted in several geometries using a variety of techniques. The transfer matrix (TM) formalism was initially used to calculate the photonic band structure of a 1D array of disk-shaped lattice sites~\cite{Deutsch1995}, similar to the situation described in Refs.~\cite{Slama2005,Slama2005a,Slama2006}. This lattice geometry recently led to the first experimental observation of photonic band gaps in an optical lattice \cite{Schilke2011,Schilke2012}.

In higher dimensions, the Bloch-Floquet model has been used to predict the occurrence of
omnidirectional photonic band gaps, for either diamond lattices~\cite{Antezza2009}, simple cubic or face-centered cubic lattices~\cite{DeshuiYu2011}. However, the generation of omnidirectional photonic bands in 3D optical lattices is technically challenging and has not yet been demonstrated in experiment.

The present paper is devoted to the theory of PBGs in one-dimensional optical lattices. It presents results obtained from a microscopic model that describes the collective scattering by a finite collection of point-like scatterers. They are compared to analytical results obtained from the TM formalism which treats scattering from perfect one-dimensional systems. In particular, we describe and discuss finite-size effects and effects due to disorder which the TM formalism cannot capture. Furthermore, we present a microscopic model treating light as a vector field, which is beyond the capability of the TM formalism. Another important part of this work concerns the possibility to manipulate the PBG by tuning the atom-radiation interaction using, for example, an external magnetic field.

\bigskip

Section~\ref{sec:theory} is devoted to the theories of one-dimensional PBG: In Section~\ref{sec:microscopic}, we present a microscopic theory where atoms are described as point-like scatterers. We show that it efficiently captures the scattering properties of the optical lattice including the one-dimensional PBG. In Section~\ref{sec:TM}, the TM formalism is introduced and derived directly from the microscopic theory. The agreements between these approaches, as well as the limitations of each, are discussed. Section~\ref{sec:vectorial} generalizes the microscopic theory by taking into account the vectorial nature of light. Section~\ref{sec:3level} completes the theoretical part with an extension of the theory of PBGs to three-level systems and a discussion of the potential role of electromagnetically
induced transparency (EIT) schemes in PBGs.

In Section~\ref{sec:experiment} we briefly present and discuss a recent and currently unique experiment which demonstrates a PBG in an optical lattice. Naturally, the experiment has peculiarities which are difficult or impossible to take into account in idealizing theoretical models, such as the TM formalism. Among them are disorder and finite-size effects. In Section~\ref{sec:FiniteSize} we discuss a number of these effects and show many of them are included in the microscopic theory.

\section{Modeling\label{sec:theory}}

Following the pioneering work of~\cite{Deutsch1995} and until
recently 1D PBGs in optical lattices have only been studied using
the TM formalism: the atomic layers are assimilated into a
dielectric medium with an infinite radius. The density is
generally assumed to be perfectly constant throughout the layer,
or at least constant transversally and stepwise longitudinally.
Although these approximations may seem to lack rigor, they have
proved to be in good agreement with the experimental measurements
of 1D PBG in optical lattices~\cite{Schilke2011}.

An alternative approach has been used to characterize the band structure of 3D optical lattices~\cite{Coevorden1996,Antezza2009}.
Working in the point-dipole approximation and decomposing the electric field into Bloch waves, Maxwell equations allow calculation of the
dispersion relation for the propagation of light within the lattice as well as the local density of states. However, the expansion into Bloch
waves is based on the assumption of an infinite lattice which is not appropriate for the optical lattices that are currently being realized
experimentally.

Recently, a microscopic theory, inspired by theories of collective scattering by a collection of point-like
scatterers~\cite{Courteille2010,Bienaime2010,Bux2010,Bienaime2011}, was introduced by~\cite{Samoylova2013} to describe PBGs.
It was shown to successfully capture the photonic properties of optical lattices, such as spontaneous emission, Bragg scattering and PBG.

This section will be devoted to the introduction of this microscopic theory (Sec.~\ref{sec:microscopic}), and its connection to TM
formalism (Sec.~\ref{sec:TM}). A formal bridge is demonstrated and numerical results of both theories are also compared with a particular
focus on PBG properties.

\subsection{Scalar microscopic theory\label{sec:microscopic}}

\subsubsection*{Model}

Here a model that describes the interaction of a
collection of static two-level atoms with a scalar wave is presented. It
accounts for the fact that each atom is affected by the radiation
of all the other atoms and so it describes multiple scattering and
collective effects. This model has been used to predict several
features of collectivity in disordered clouds, such as
the modification of the radiation pressure
force~\cite{Courteille2010}, Mie scattering and
resonances~\cite{Bender2010,Bachelard2012} or the collective Lamb
shift~\cite{Bienaime2011}. While the model is formally the same
for ordered atoms, as first discussed in
Ref.~\cite{Samoylova2013}, the scattering properties of
optical lattices are very different.

The atomic cloud is described as a system of two-level ($g$ and
$e$) atoms, with resonant frequency $\omega_a$ and position
$\mathbf{r}_j$, which are driven by a uniform laser beam with
electric field amplitude $E_0$, frequency $\omega_0$ and wave
vector $\mathbf{k}_0=(\omega_0/c)\mathbf{\hat e}_z$. The
atom-laser interaction is described by the Hamiltonian:
\begin{eqnarray}\label{eq:H0}
\hat{H}&=&\frac{\hbar\Omega_0}{2}\sum_{j=1}^N\left[\hat\sigma_j
e^{i(\Delta_0 t- \mathbf{k}_0\cdot
\mathbf{r}_j)}+\textrm{h.c.}\right]\\ \nonumber &+&
\hbar\sum_{j=1}^N\sum_{\mathbf{k}}g_k\left(\hat\sigma_j
e^{-i\omega_a t} +\hat\sigma_j^\dagger e^{i\omega_a t}\right)
\left[\hat a_{\mathbf{k}}^\dagger e^{i(\omega_k t- \mathbf{k}\cdot
\mathbf{r}_j)}+\hat a_{\mathbf{k}} e^{-i(\omega_k t-
\mathbf{k}\cdot \mathbf{r}_j)}\right],
\end{eqnarray}
where $\Omega_0=d E_0/\hbar$ is the Rabi frequency of the incident
laser field and $\Delta_0=\omega_0-\omega_a$ is the detuning
between the laser and the atomic transition. For simplicity scalar light is considered in ~(\ref{eq:H0}), with a vectorial form of the model being discussed later in
Sec.~\ref{sec:vectorial}. The case of three-level atoms is
discussed in Sec.~\ref{sec:3level}.

In (\ref{eq:H0}), $\hat\sigma_j=|g_j\rangle\langle e_j|$ is the
lowering operator for the $j$th atom, $\hat a_{\mathbf{k}}$ is the
photon annihilation operator and
$g_k=(d^2\omega_a^2/2\hbar\epsilon_0 \omega_kV_\nu)^{1/2}$ is the
single-photon Rabi frequency, where $d$ is the electric-dipole
transition matrix element and $V_\nu$ is the photon volume. The
special case of a low-intensity laser, where a single photon from
mode $\mathbf{k}$ can be assumed to be present in the system, was
extensively investigated in
Refs.~\cite{Friedberg1973,Scully2006,Svidzinsky2008}. In this case
the system (atoms+photons) is described by a state of the
form~\cite{Svidzinsky2010}:
\begin{eqnarray}\label{state}
    |\Psi\rangle&=&\alpha(t)|g_1\dots g_N\rangle |0\rangle_{\mathbf{k}}+e^{-i\Delta_0 t}\sum_{j=1}^N
    \beta_j(t)|g_1\ldots e_j\ldots g_N\rangle|0\rangle_{\mathbf{k}}\\
    &+& \sum_{\mathbf{k}}\gamma_{\mathbf{k}}(t)|g_1\dots g_N\rangle
    |1\rangle_{\mathbf{k}}+\sum_{\mathbf{k}}\sum_{j\neq m}
    \epsilon_{j,m,\mathbf{k}}(t)|g_1\ldots e_j\ldots e_m\ldots
    g_N\rangle|1\rangle_{\mathbf{k}}.\nonumber
\end{eqnarray}
The first term in ~\eqref{state} corresponds to the initial
ground state without photons, the second term is the sum over the
states where a single atom has been excited by the classical
field, and the third term corresponds to the atoms that returned
to the ground state having emitted a photon in the mode
$\mathbf{k}$. Hence, the square modulus of $\alpha$, $\beta_j$ and $\gamma_\mathbf{k}$ represent respectively the probability that the is no photon in the system and no excited atom, the probability that atom $j$ is excited (and no photon), and the probability to have a photon in mode $\mathbf{k}$ (and all atoms in the ground state). Finally, $|\epsilon_{j,m,\mathbf{k}}|^2$ describes the probability of having two excited atoms and one virtual photon in mode $\mathbf{k}$ with
`negative' energy. This last term of \eqref{state} is due to the counter-rotating terms in the
Hamiltonian (\ref{eq:H0}) and disappears when the rotating wave
approximation is made. The scattering process using the latter
approximation was studied in several
references~\cite{Mazets2007,Courteille2010,Bienaime2010,Skipetrov2011,Bachelard2011},
but the importance of counter-rotating terms during the relaxation
process was pointed out in Ref.~\cite{Manassah2012a} .

The dynamics of each component of the state \eqref{state} are deduced from the Schr\"odinger equation:
\begin{equation}
\frac{\partial|\Psi(t)\rangle}{\partial t} =-\frac{i}{\hbar}\hat{H}|\Psi(t)\rangle.
\end{equation}
Hence, the Hamiltonian~\eqref{eq:H0} leads to the following set of differential equations:
\begin{align}
    \dot{\alpha}(t)  =&-\tfrac{i}{2}\Omega_0e^{i\Delta_0t}\sum_{j=1}^N\beta_j(t)e^{-i\mathbf{k}_0\cdot\mathbf{r}_j}~,\label{EqAmplitude1}\\
    \dot{\beta}_j(t) =&i\Delta_0\beta_j-\tfrac{i}{2}\Omega_0\alpha(t)e^{-i\Delta_0t+i\mathbf{k}_0\cdot\mathbf{r}_j}
    -i\sum_{\mathbf{k}}g_k\gamma_{\mathbf{k}}(t)e^{-i(\omega_k-\omega_0)t+i\mathbf{k}\cdot \mathbf{r}_j}\nonumber\\
    & -i\sum_{\mathbf{k}}g_k e^{-i(\omega_k+\omega_a-\Delta_0)t}\sum_{m\neq j}
    \epsilon_{j,m,\mathbf{k}}e^{i\mathbf{k}\cdot\mathbf{r}_j} ~,\label{EqAmplitude2}\\
    \dot{\gamma}_{\mathbf{k}}(t)
    =&-ig_ke^{i(\omega_k-\omega_0)t}\sum_{j=1}^N\beta_j(t)e^{-i\mathbf{k}\cdot\mathbf{r}_j}~.\label{EqAmplitude3}\\
    \dot{\epsilon}_{j,m,\mathbf{k}}(t) =&
   -ig_{k}e^{i(\omega_k+\omega_a-\Delta_0)t}\left[
  e^{-i\mathbf{k}\cdot \mathbf{r}_j}\beta_m+e^{-i\mathbf{k}\cdot \mathbf{r}_m}\beta_j\right]. \label{EqAmplitude4}
\end{align}
Then integrating Eqs.~(\ref{EqAmplitude3}) and
(\ref{EqAmplitude4}) over time with $\gamma_{\mathbf{k}}(0)=0$ and
$\epsilon_{j,m,\mathbf{k}}(0)=0$ and substituting them in to
~\eqref{EqAmplitude2}, we obtain $N$ coupled equations
describing the dynamics of the atomic dipoles:
\begin{align}\label{EqEvolution}
    \dot{\beta}_j(t)= &i\Delta_0\beta_j -\tfrac{i}{2}\Omega_0\alpha e^{i\mathbf{k}_0\cdot\mathbf{r}_j}\\
    & -\sum_{\mathbf{k}}g_k^2 \sum_{m=1}^Ne^{i\mathbf{k}\cdot(\mathbf{r}_j-\mathbf{r}_m)}\int_0^{t}
    e^{-i(\omega_k-\omega_0)(t-t')}\beta_m(t')dt'\nonumber\\
    &-\sum_{\mathbf{k}}g_{k}^2\int_0^t
dt'e^{i(\Delta_0-\omega_k-\omega_a)(t-t')}\nonumber\\
&\times  \left[\sum_{m\neq j}
e^{-i\mathbf{k}\cdot(\mathbf{r}_j-\mathbf{r}_m)}\beta_m(t')+(N-1)\beta_j(t')
\right].
\end{align}
The interaction with the vacuum field yields diagonal terms with
$m=j$ whose real part corresponds to the single-atom decay term
and imaginary part to the  self-energy shift, and off-diagonal
terms with $m\neq j$ which correspond to the atom-atom interaction
mediated by the photon. By separating the two contributions, we
can write:
\begin{align}\label{EqEvolution2}
    \dot{\beta}_j(t)= &i\Delta_0\beta_j - \tfrac{i}{2}\Omega_0\alpha e^{i\mathbf{k}_0\cdot\mathbf{r}_j}\nonumber\\
    &-\sum_{\mathbf{k}}g_{k}^2\int_0^t
d\tau
\left[e^{-i(\omega_k-\omega_0)\tau}+e^{i(\Delta_0-\omega_k-\omega_a)\tau}(N-1)
\right]\beta_j(t-\tau)\nonumber\\
& -\sum_{\mathbf{k}}g_k^2 \sum_{m\neq
j}e^{i\mathbf{k}\cdot(\mathbf{r}_j-\mathbf{r}_m)}\nonumber\\
&\times \int_0^{t}d\tau
    \left[e^{-i(\omega_k-\omega_0)\tau}
    + e^{i(\Delta_0-\omega_k-\omega_a)\tau}\right]\beta_m(t-\tau).
\end{align}
We assume a smooth density of modes, so the discrete sum
$\sum_\mathbf{k}$ can be replaced by the volume integral
$[V_\nu/(2\pi)^3]\int d\mathbf{k}$. In the linear regime
$\alpha\approx 1$ and in the Markov approximation, valid if the
decay time is larger than the  photon time-of-flight through the
atomic cloud, in the integrals of
(\ref{EqEvolution2}) we can replace $\beta_j(t-\tau)$ with $\beta_j(t)$ and
extend the upper integration limit to infinity, so that
(\ref{EqEvolution2}) is approximated by
\begin{align}\label{EqEvolution3}
    \dot{\beta}_j(t)= & i\Delta_0\beta_j - \tfrac{i}{2}\Omega_0 e^{i\mathbf{k}_0\cdot\mathbf{r}_j}\nonumber\\
    &-\frac{V_\nu}{(2\pi)^3}\int d\mathbf{k}g_{k}^2\int_0^\infty
d\tau
\left[e^{-i(\omega_k-\omega_0)\tau}+e^{-i(\omega_k+2\omega_a-\omega_0)\tau}(N-1)
\right]\beta_j(t)\nonumber\\
& -\frac{V_\nu}{(2\pi)^3}\int d\mathbf{k}g_k^2 \sum_{m\neq
j}e^{i\mathbf{k}\cdot(\mathbf{r}_j-\mathbf{r}_m)}\nonumber\\
&\times \int_0^{\infty}d\tau
    \left[e^{-i(\omega_k-\omega_0)\tau}
    + e^{-i(\omega_k+2\omega_a-\omega_0)\tau}\right]\beta_m(t).
\end{align}
The imaginary part of the self-field contribution (third term of
the right hand side of (\ref{EqEvolution3})) consists of the
self-energy shift of the atom in the excited state plus the
self-energy contribution of the atom in the ground state. The
effect is an adjustment to the transition frequency $\omega_a$, which is assumed to already include the shift. It can not be
computed realistically using our model, which treats the atoms as
point particles, and is disregarded in the present approach.
After performing the integration over $\tau$ the real part of the self-field contribution yields:
\begin{eqnarray}\label{gammaspo}
  \frac{V_\nu}{2\pi}\int_0^\infty dk
    k^2g_{k}^2\left[\delta(\omega_k-\omega_0)+(N-1)\delta(\omega_k+2\omega_a-\omega_0)\right]
    =\frac{\Gamma}{2},
\end{eqnarray}
where $\omega_k=ck$ and $\Gamma=d^2\omega_0^3/(2\pi
c^3\hbar\epsilon_0)$ is the single-atom {\it spontaneous} decay
rate in the radiation scalar theory. The last term in
(\ref{gammaspo}) arises from the counter-rotating wave terms of
the Hamiltonian (\ref{eq:H0}) and does not contribute as it
corresponds to a negative photon energy
$\omega_k\approx-\omega_a$. Using (\ref{gammaspo}) in
(\ref{EqEvolution3}) we obtain:
\begin{align}\label{EqEvolution4}
    \dot{\beta}_j(t)= & \left(i\Delta_0-\frac{\Gamma}{2}\right)\beta_j- \tfrac{i}{2}\Omega_0 e^{i\mathbf{k}_0\cdot\mathbf{r}_j}
 -\frac{V_\nu}{(2\pi)^3}\int d\mathbf{k}g_k^2 \sum_{m\neq
j}e^{i\mathbf{k}\cdot(\mathbf{r}_j-\mathbf{r}_m)}\nonumber\\
&\times \int_0^{\infty}d\tau
    \left[e^{-i(\omega_k-\omega_0)\tau}
    + e^{-i(\omega_k+2\omega_a-\omega_0)\tau}\right]\beta_m(t).
\end{align}
The last term on the right hand side of (\ref{EqEvolution4})
can be calculated as follows:
\begin{align}\label{integr}
& \frac{V_\nu}{(2\pi)^3}\int d\mathbf{k}g_k^2
e^{i\mathbf{k}\cdot(\mathbf{r}_j-\mathbf{r}_m)}
\int_0^{\infty}d\tau
    \left[e^{-i(\omega_k-\omega_0)\tau}
    + e^{-i(\omega_k+2\omega_a-\omega_0)\tau}\right]\nonumber\\
    &=\frac{c\Gamma}{\pi k_0}\int_0^{\infty}d\tau \cos(\omega_0\tau)
    \int_0^\infty dk k
\frac{\sin(kr_{jm})}{kr_{jm}}e^{-ick\tau},
\end{align}
where $r_{jm}=|\mathbf{r}_j-\mathbf{r}_m|$ and we assumed
$\omega_0\approx\omega_a$. We use the integral expression
\[
\int_0^\infty dk k \frac{\sin(kr)}{kr}e^{-ick\tau}= \frac{1}{2cr}
\left[ \frac{1}{\tau+r/c-i\epsilon}-\frac{1}{\tau-r/c-i\epsilon}
\right]
\]
where $\epsilon\rightarrow 0^+$. Changing the sign of the integration
variable $\tau$ in the first term, (\ref{integr}) becomes
\begin{align}\label{intint}
    &-\frac{\Gamma}{2\pi k_0r_{jm}}\int_{-\infty}^{\infty}d\tau\frac{\cos(\omega_0\tau)}{\tau-r_{jm}/c+i\epsilon}
    =\frac{\Gamma}{2}\frac{\exp(ik_0r_{jm})}{ik_0r_{jm}}.
\end{align}
Finally, the scattering problem reduces to the set of differential
equations~\cite{Scully2009,Bachelard2011,Bienaime2011}:
 \begin{eqnarray}\label{eqbetaj}
    \dot\beta_j&=&\left(i\Delta_0-\frac{\Gamma}{2}\right)\beta_j- i\frac{\Omega_0}{2}e^{i \mathbf{k}_0\cdot
    \mathbf{r}_j}-\frac{\Gamma}{2}\sum_{m\neq j}
    \frac{\exp(ik_0|\mathbf{r}_j-\mathbf{r}_m|)}{ik_0|\mathbf{r}_j-\mathbf{r}_m|}\beta_m.
\end{eqnarray}
The kernel in the last term of (\ref{eqbetaj}) has a real
component, $-(\Gamma/2)\sum_{m\neq
j}\text{sinc}(k_0|\mathbf{r}_j-\mathbf{r}_m|)$, describing the
{\it collective} atomic decay, and an imaginary component,
$i(\Gamma/2)\sum_{m\neq
j}\cos(k_0|\mathbf{r}_j-\mathbf{r}_m|)/(k_0|\mathbf{r}_j-\mathbf{r}_m|)$,
which contains the collective Lamb shift due to short range
interactions between atoms induced by the electromagnetic
field~\cite{Scully2009,Scully2010,Rohlsberger2010,Keaveney2012}.

In what follows we will focus on the stationary state of~(\ref{eqbetaj}), which reads
 \begin{equation}\label{eqbetajstat}
    \left(2\Delta_0+i\Gamma\right)\beta_j= \Omega_0 e^{i \mathbf{k}_0\cdot
    \mathbf{r}_j}-\Gamma\sum_{m\neq j}
    \frac{\exp(ik_0|\mathbf{r}_j-\mathbf{r}_m|)}{k_0|\mathbf{r}_j-\mathbf{r}_m|}\beta_m.
\end{equation}

Practically, the value of the atomic dipoles $\beta_j$ in the stationary regime is evaluated numerically by inverting the linear problem \eqref{eqbetajstat}, since it is easily cast in the form $M\vec{\beta}=e^{i\mathbf{k}_0\cdot\vec{\mathbf{r}}}$, where $\vec{\beta}$ and $\vec{\mathbf{r}}$ refer to the vectors of $\beta_j$ and $\mathbf{r}_j$. From the $\beta_j$, the observables described in the next sections are easily calculated.

While we adopted a quantum, single-photon treatment to derive a
description of the collective scattering a classical interpretation
is also possible. In fact, ~(\ref{eqbetaj}) also describes
 the dynamics of atomic dipoles driven by the {\it total}
electric field, the sum of the incident field and of the field
radiated by the other dipoles. Indeed, the last term of
~(\eqref{eqbetaj}) gives the emission of spherical waves by
the latter, as described by Huygens principle. Furthermore, model
\eqref{eqbetaj} has also been derived through another classical approach which
treats the two-level atoms as weakly excited classical harmonic
oscillators~\cite{Svidzinsky2010,Prasad2010}.

\subsubsection*{Radiation field}

The scattered radiation is provided by the positive-frequency component of the electric field
\begin{equation}\label{Es:a}
E_S(\mathbf{r},t)=\sum_{\mathbf{k}}{\cal E}_{k}\gamma_{\mathbf{k}}(t)e^{i\mathbf{k}\cdot \mathbf{r}-i\omega_k t},
\end{equation}
where ${\cal E}_k=(\hbar\omega_k/2\epsilon_0 V_\nu)^{1/2}$ is the
single-photon electric field. We then integrate
~\eqref{EqAmplitude3} over time, with
$\gamma_{\mathbf{k}}(0)=0$, insert it in to ~\eqref{Es:a} and
obtain
\begin{equation}\label{Es:1}
    E_S(\mathbf{r},t)=-\frac{iV_\nu}{8\pi^3}e^{-i\omega_0 t}\sum_{j=1}^N
    \int d\mathbf{k}\,{\cal E}_{k}g_k\int_0^t dt'e^{-i(\omega_k-\omega_0)t'} e^{i\mathbf{k}\cdot(\mathbf{r}-\mathbf{r}_j)}\beta_j(t-t').
\end{equation}
Taking $d\mathbf{k}=k^2 dk\,\sin\theta d\phi\,d\theta$ and
perfoming the integrations over $\theta$ and $\phi$,
(\ref{Es:1}) becomes
\begin{equation}\label{Es:2}
    E_S(\mathbf{r},t)=-\frac{id\omega_0}{4\pi^2\epsilon_0}e^{-i\omega_0 t}\sum_{j=1}^N
    \int_0^t dt'\beta_j(t-t')
    \int_0^\infty dk\,k\,\frac{\sin(k|\mathbf{r}-\mathbf{r}_j|)}{|\mathbf{r}-\mathbf{r}_j|}
    e^{-i(\omega_k-\omega_0)t'}.
\end{equation}
The scattered intensity will be centered about the
incidence laser frequency $\omega_0$. The quantity $\omega_k=ck$
varies little around $\omega_k=\omega_0$ for which the time
integral in (\ref{Es:2}) is not negligible. We can therefore
replace $k$ by $\omega_0/c$ and the lower limit in the $k$
integration by $-\infty$. The integral
\begin{equation}\label{int:E}
    \int_{-\infty}^\infty
    dk\sin(kR)e^{-ic(k-k_0)t'}=\frac{\pi}{ic}
    \left[e^{ik_0R}\delta(t-R/c)-e^{-ik_0R}\delta(t+R/c)\right]
\end{equation}
yields
\begin{equation}\label{Es:3}
    E_S(\mathbf{r},t)= -\frac{dk_0^2}{4\pi\epsilon_0}\sum_{j=1}^N
    \frac{e^{ik_0 |\mathbf{r}-\mathbf{r}_j|}}{|\mathbf{r}-\mathbf{r}_j|}
    \beta_j(t-|\mathbf{r}-\mathbf{r}_j|/c).
\end{equation}
The delay in $\beta_j$ can be neglected, as previously assumed in \eqref{EqEvolution2}, if the photon
time-of-flight $|\mathbf{r}-\mathbf{r}_j|/c$ is much smaller than
the characteristic time during which the atomic variables change
appreciably, obtaining
\begin{equation}\label{Es:3bis}
    E_S(\mathbf{r},t)= -\frac{dk_0^2}{4\pi\epsilon_0}\sum_{j=1}^N
    \frac{e^{ik_0 |\mathbf{r}-\mathbf{r}_j|}}{|\mathbf{r}-\mathbf{r}_j|}
    \beta_j(t).
\end{equation}
At distances $r$ much larger than the cloud, a far-field
expression can be derived. Using $|\mathbf{r}-\mathbf{r}_j|\approx
r-\hat{ \mathbf{n}}\cdot\mathbf{r}_j$, where
$\hat{\mathbf{n}}=\mathbf{r}/r$, ~\eqref{Es:3bis} turns
into
\begin{equation}\label{Es}
    E_S(\mathbf{k})\approx
    -\frac{dk_0^2}{4\pi\epsilon_0}\frac{e^{ik_0r}}{r}\sum_{j=1}^N
    \beta_j(t) e^{-i\mathbf{k}\cdot\mathbf{r}_j},
\end{equation}
where $\mathbf{k}=k_0\hat{\mathbf{n}}$. Notice that with
~\eqref{Es:3bis} we recover the field present in
~\eqref{eqbetajstat} apart from the self-contribution of
$j$th-atom to the field. Thus, the excitation of the atomic dipole
in ~\eqref{eqbetajstat} can be rewritten as
\begin{equation}
\beta_j=\frac{d}{\hbar\left(\Delta_0+\Gamma/2\right)}E_{tot\setminus
j}(\mathbf{r}_j),
\end{equation}
where $E_{tot\setminus
j}(\mathbf{r}_j)=(E_0/2)\exp(i\mathbf{k}_0\cdot
\mathbf{r}_j)+E_S(\mathbf{r}_j)-E_{\mathrm{self}}(\mathbf{r}_j)$
is the {\it total} electric field minus the atom
self-contribution. Hence, ~\eqref{eqbetajstat} describes the
light response of a set of point-like dielectric particles (in the
linear optics regime), where the self-field contributions are
treated separately to avoid singularities.

Measurements of the atomic density distribution after a time of free expansion yield the atomic momentum distribution after interaction with the light beam. The momentum distribution, which is modified by the recoil imparted to the atomic cloud due to the radiation pressure force, contains important information on the light scattering in
disordered
systems~\cite{Courteille2010,Bachelard2012a,Piovella2013}. The
strong relationships between scattered light and radiation
pressure force are a manifestation of collective effects induced
by scattered photons and described in the present
model~\cite{Chabe2013}. Until now we assumed free atoms. However, in optical lattices the laser beams used to create
the lattice potential also generate a strong confining force localizing the atoms within a regime smaller than the Lamb Dicke limit, so that
the simple link between scattered light and radiation pressure
force is broken. Other effects which take into account the atomic trapping need to
be included in order to access more exhaustive information revealed by
time-of-flight measurements.

\subsubsection*{Bragg scattering}

Bragg scattering is the reflection of light by a periodic structure due to constructive interference. This phenomenon has been used to probe atomic structure~\cite{Wollan1932}, molecular dynamics~\cite{Doucet1987}, optical lattices~\cite{Birkl1995,Weidemueller1995,Weidemueller1998}, and photonic bands in photonic crystals~\cite{Koenderink2003}.

In a one-dimensional structure of period $d$ a wave with wavelength $\lambda_0$ is reflected provided its incident angle $\theta_0$ (with respect to the structure axis) satisfies the Bragg condition:
\begin{equation}
\sin\theta_0=n\frac{\lambda_0}{2d},\label{eq:BraggAngle}
\end{equation}
where $n$ is an integer. The condition \eqref{eq:BraggAngle} is actually that required to provide constructive interference from the scatterers. In the case of an atomic lattice dense enough to modify the phase of the propagating wave, a correction is necessary to account for the medium effective index $m$~\cite{Deutsch1995}, and the Bragg condition turns into
\begin{equation}
\sin\theta_0=n\frac{\lambda_0}{2md}.\label{eq:AtomicBraggAngle}
\end{equation}

Model \eqref{eqbetajstat} naturally describes Bragg scattering since the radiation field \eqref{Es:3} takes into account the interference from the atomic scatterers. Yet the collective term in \eqref{eqbetajstat} which describes the radiation from the neighbour is unimportant as far as Bragg scattering is concerned. Let us consider the radiation of the lattice without the collectivity. Then ~\eqref{eqbetajstat} simply turns into
\begin{equation}\label{eq:BetajNonColl}
    \beta_j= \Omega_0\frac{e^{i \mathbf{k}_0\cdot
    \mathbf{r}_j}}{\left(2\Delta_0+i\Gamma\right)}.
\end{equation}
Each atomic dipole follows the laser only, from which follows a state that is similar to the mean-field ansatz called the timed Dicke-state~\cite{Scully2006}. The many-body aspect of the problem only appears as the field radiated by the atoms interferes in ~\eqref{Es:3}, but this interference is sufficient to modify the structure factor of the atomic lattice and reflect the incoming wave. This effect is described by Rayleigh scattering, where the synchronization of the atomic dipoles by coherent light results in long-range correlations and in an emission of light in the Bragg directions for large clouds. However, here the lattice period constrains the light emission to specific directions.

\begin{figure}[ht]
{\psfig{file=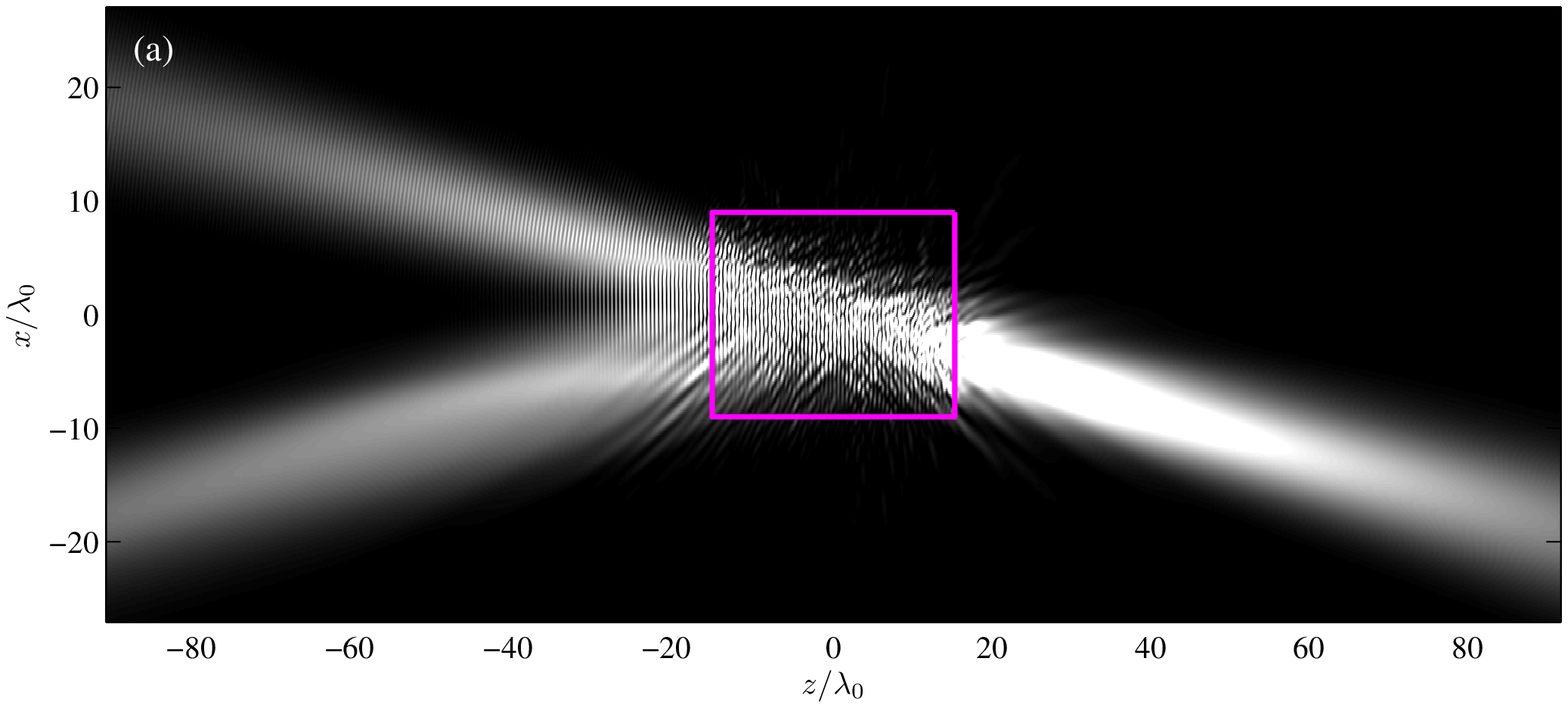,width=10cm}
\\ \psfig{file=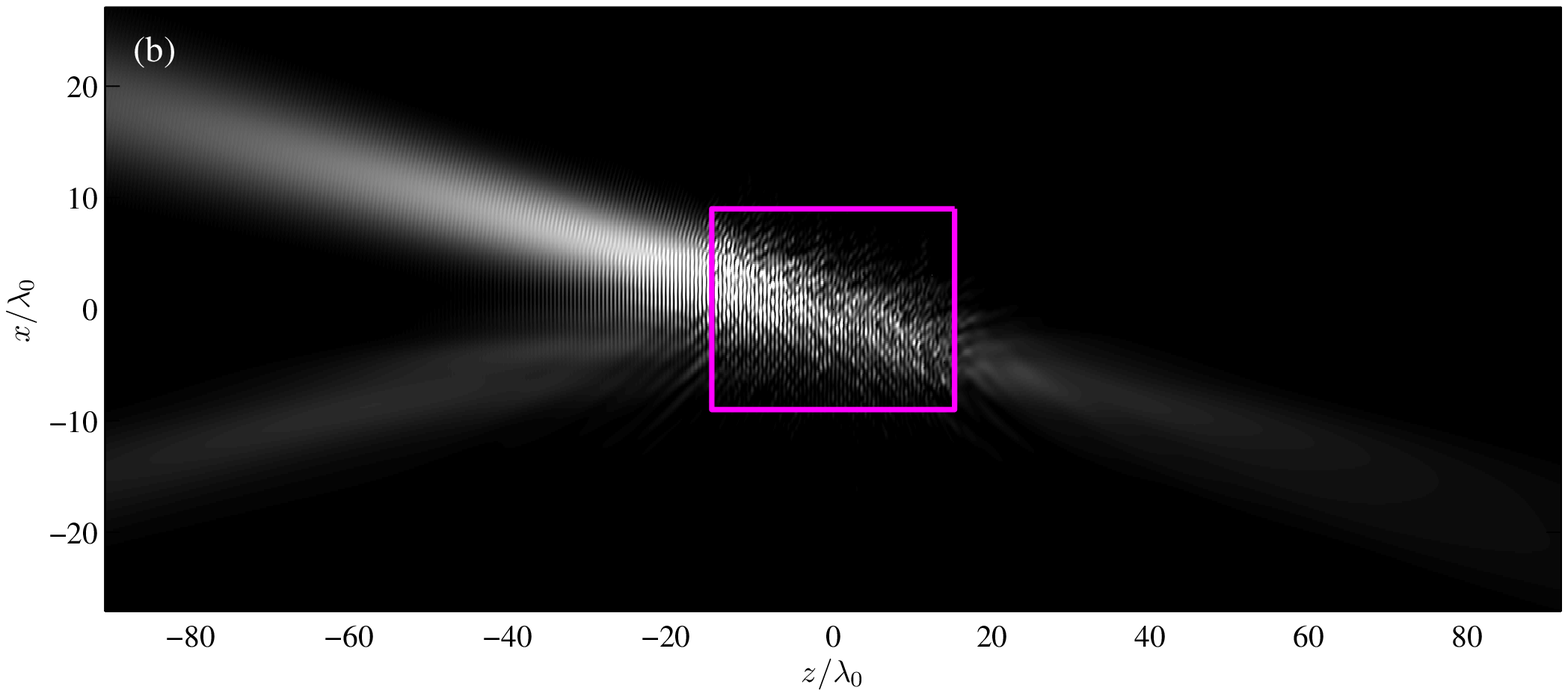,width=10cm}}
\caption{Bragg reflection of a Gaussian beam by a one-dimensional optical lattice for models \eqref{eq:BetajNonColl} neglecting collective (multiple) scattering [top] and \eqref{eqbetajstat} including collective scattering [bottom]. The laser beam has a waist $4.5\lambda_0$, a detuning $\Delta_0=\Gamma$, and arrives from the top left with angle $\theta_0=0.2$rad. The lattice is composed of $5000$ atoms spread over $60$ disks of radius $9\lambda_0$ and thickness $0.04\lambda_0$, separated by $d=\lambda_0/2\cos\theta_0$. The rectangles mark the limit of the atomic lattice.}
\label{fig:BraggScattering}
\end{figure}

Bragg scattering of a Gaussian beam using models \eqref{eq:BetajNonColl} and \eqref{eqbetajstat} is displayed in Fig.~\ref{fig:BraggScattering}.  Despite both describing the light reflection, model \eqref{eq:BetajNonColl} does not conserve energy, nor is it able to account for the progressive attenuation of the light in the lattice as the photons are reflected. This highlights the necessity of including collective terms, such as those in \eqref{eqbetajstat}, when investigating PBGs.
We point out that the study of photonic band gaps on a finite cloud requires a finite beam, thus the Gaussian beam used above. Indeed if one considers the scattering of a plane-wave  on a finite lattice, the effect of the cloud in the far-field limit becomes negligible because of diffraction.

\subsubsection*{One-dimensional photonic band gap}

A PBG corresponds to a range of light frequencies which cannot propagate through the medium, and instead are reflected. As discussed above accounting for multiple scattering, and thus collective effects, is necessary to describe the progressive reflection of the light as it penetrates the lattice. We here show how one-dimensional PBGs are described by the collective model \eqref{eqbetajstat}.

\begin{figure}[ht]
{\psfig{file=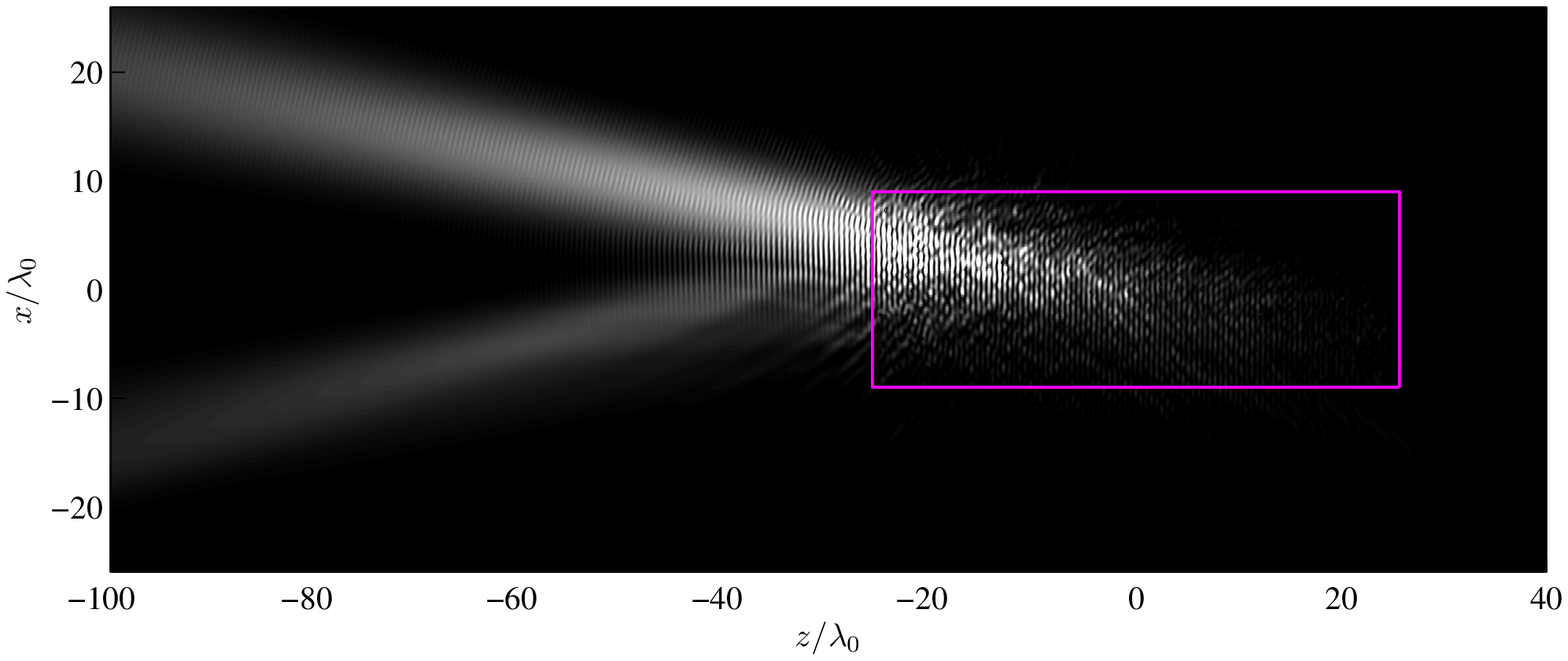,width=10cm}\\ \psfig{file=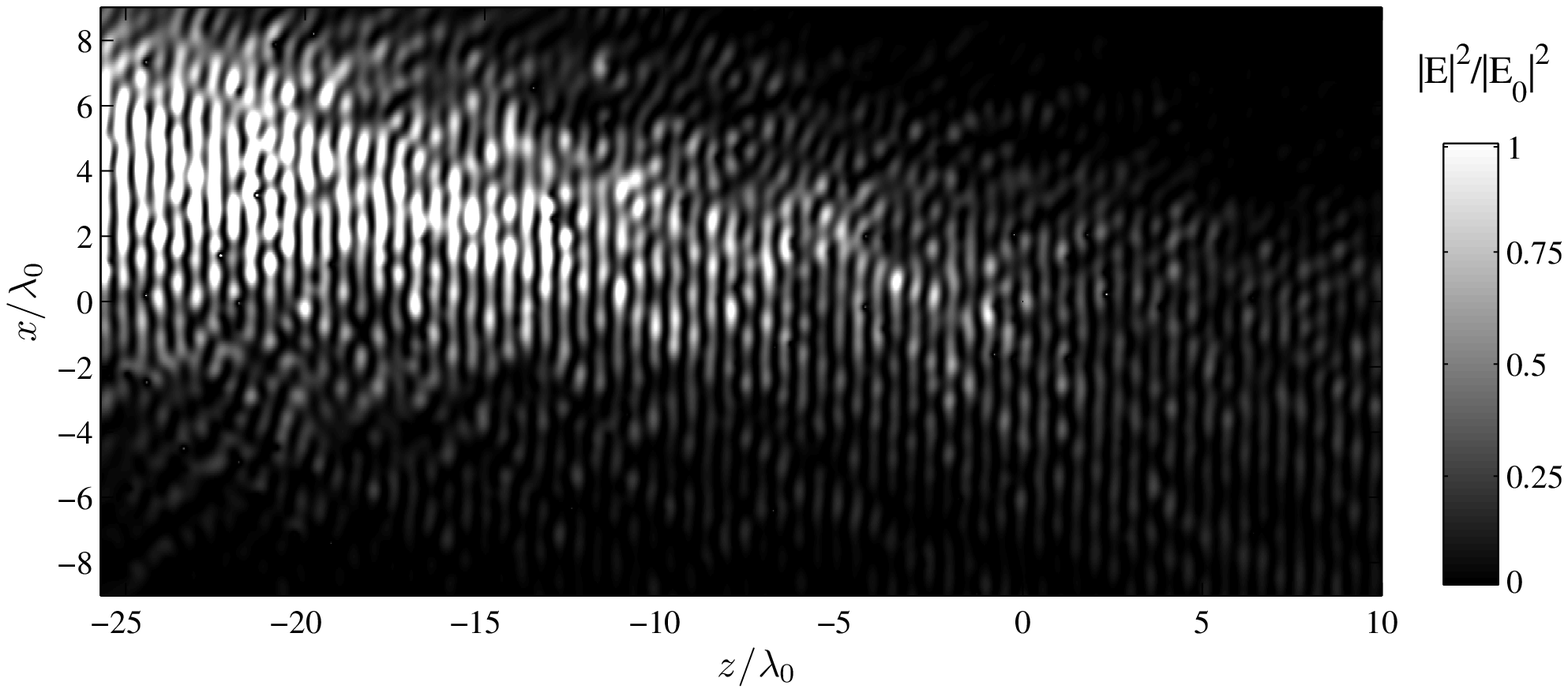,width=10cm}}
\caption{Top: Intensity of the light in the $y=0$ plane as it enters a one-dimensional optical lattice and its reflection.
The rectangle marks the limit of the atomic structure. Bottom: Zoom of the leftmost region of the atomic lattice. The luminous grains correspond to the strong
field radiated by the atoms close to the $y=0$ plane. The simulations are realized for $N=9000$ atoms randomly distributed over $N_d=100$ layers of
thickness $a=0.06\lambda_0$ and radius $R=9\lambda_0$, the distance between the atomic disks is $d=0.508\lambda_0$ with $\lambda_0$ being the
resonance wavelength. The input Gaussian beam has waist $4.5\lambda_0$ and power $100~$mW with a detuning $\Delta_0=\Gamma$ and is incident at an angle $\theta_0=0.2~$rad with respect to the lattice axis.}
\label{fig:PBG}
\end{figure}

As shown in Fig.~\ref{fig:PBG}, the microscopic model \eqref{eqbetajstat} provides a good description of the reflection of the light. Aside from a spontaneous emission contribution that is emitted in all directions, most of the light is reflected and almost no light propagates in transmission behind the lattice. Note that only an infinite lattice is actually able to completely stop the light.

The capacity for a given wavelength to propagate in a medium is quantified by the local density of states (LDOS). In the case of one-dimensional systems the LDOS at the center of the lattice can be conveniently calculated using the complex
reflection coefficients $r_{-,+}$ corresponding to the reflection from the two halves of the lattice, i.e., from the lattice beginning to the center, and from the center to the end~\cite{Boedecker2003}:
\begin{equation}\label{eq:LDOS}
    LDOS=\text{Re}\left(\frac{2+r_-+r_+}{1-r_-r_+}-1\right).
\end{equation}
The complex reflection coefficient $r_-=\sqrt{R_-}e^{i\phi}$ is
computed numerically using the reflectivity $R_-$ of the first
semi-lattice, i.e., the ratio of the reflected to the incident
power, and the phase $\phi$ of the wave reflected at the origin of
the lattice. 

\begin{figure}[ht]
{\psfig{file=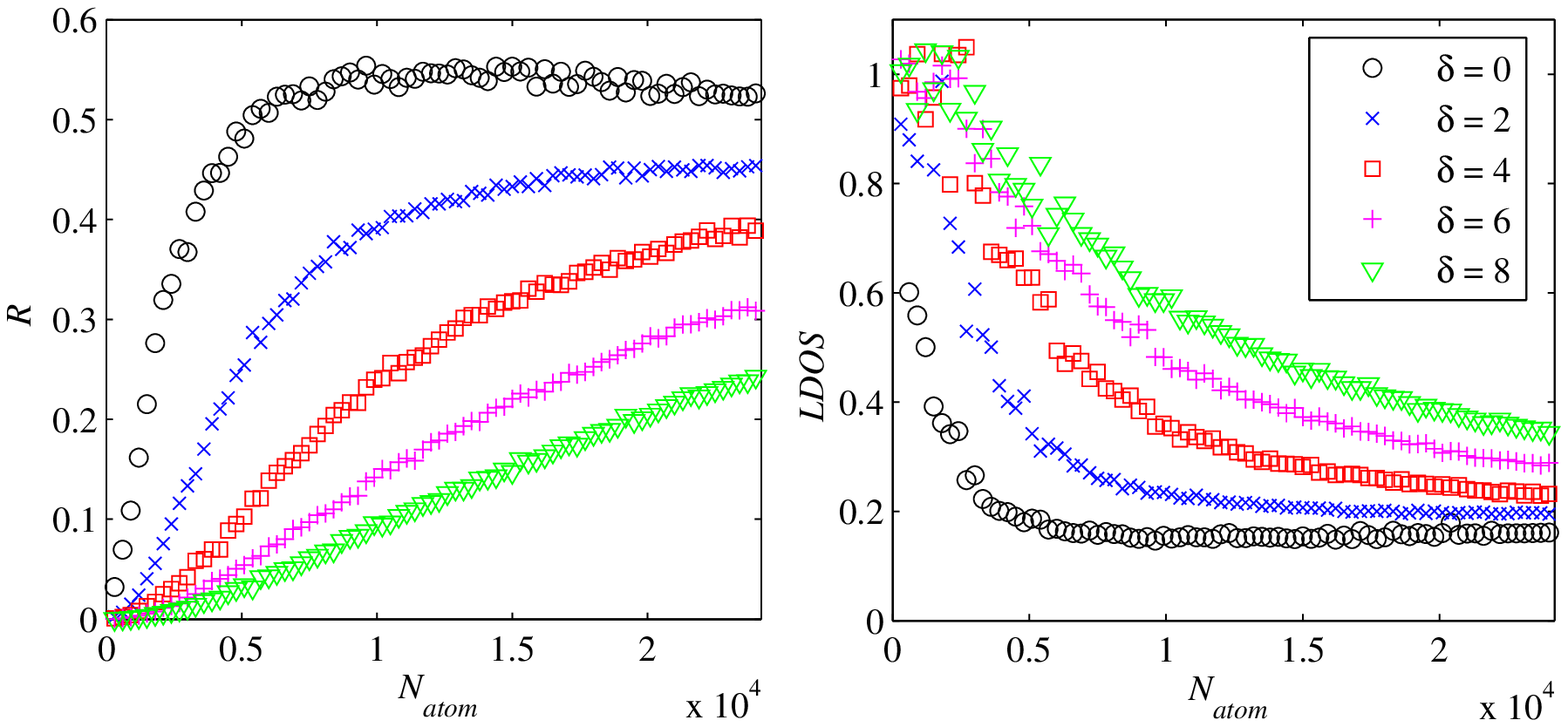,width=12cm}}
\caption{Reflection coefficient $R$ (left) and LDOS (right) as a function of the number of atoms in the lattice, for detunings $\delta=0$, $2$, $4$, $6$ and $8$. The one-dimensional lattice is composed of $N_d$ disks with $60$ atoms each, with spacing $d=\lambda_0/2$, disk radius $9\lambda_0$ and thickness $0.04\lambda_0$. The laser waist is $4.5\lambda_0$.}
\label{fig:R-LDOS-N}
\end{figure}

The quantity of light reflected by the lattice naturally increases with the number of atomic layers, as can be observed in Fig.~\ref{fig:R-LDOS-N}. Furthermore, the larger the detuning the weaker the interaction between the light and each layer, and therefore the longer the lattice needs be to efficiently reflect the incoming light. In particular, larger detunings allow for a reduced spontaneous emission since the imaginary part of the atomic polarizability is reduced, yet this requires lattice lengths beyond what can be feasibly be simulated numerically using the model \eqref{eqbetajstat}. Consequently in what follows we focus on resonant or near-resonant light.

\begin{figure}[ht]
{\psfig{file=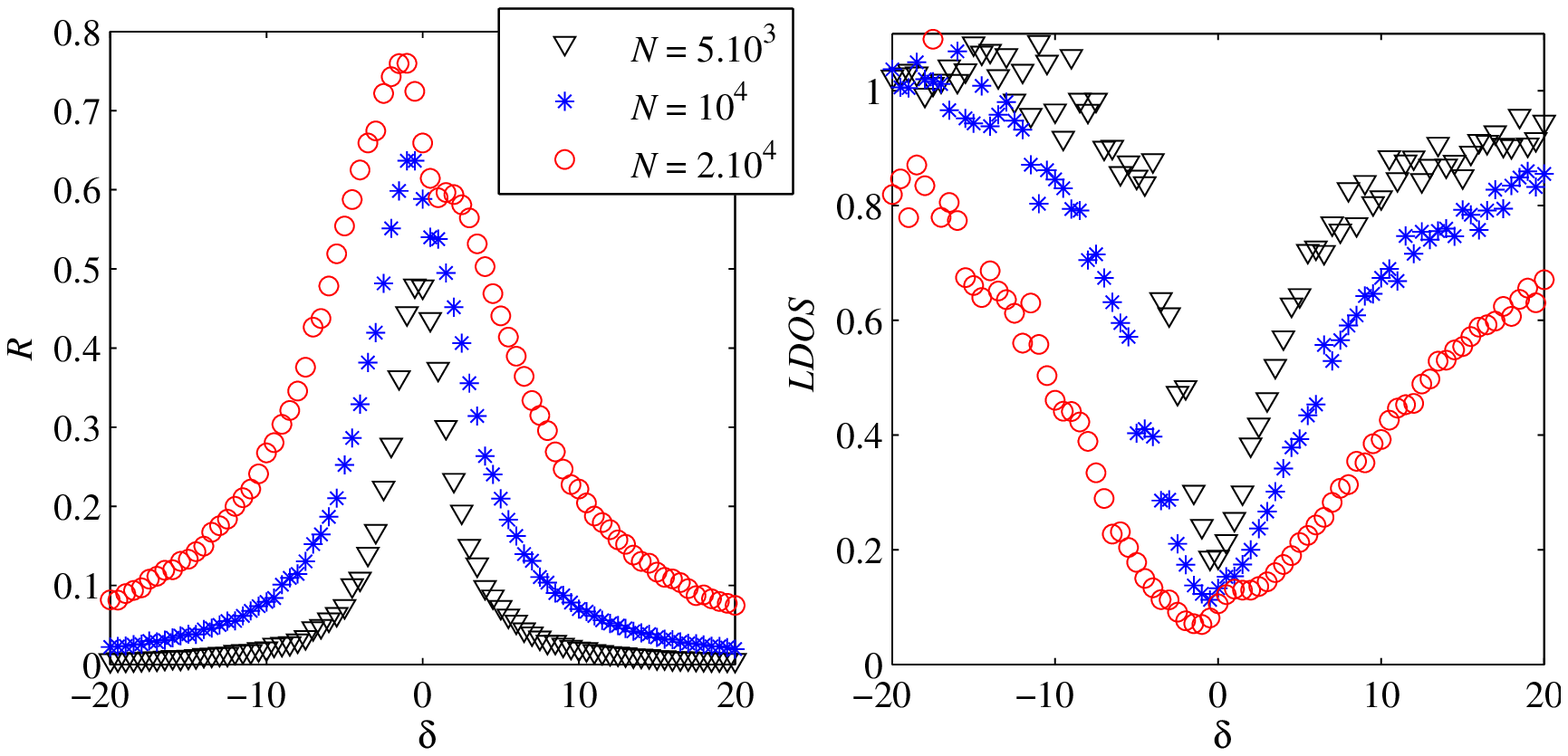,width=12cm}}
\caption{Reflection coefficient $R$ (left) and LDOS (right) as a function of the detuning $\delta$ for atom numbers $N=5.10^3$, $10^4$ and $2.10^4$. The one-dimensional lattice is composed of $N_d=100$ disks, with spacing $d=\lambda_0/2$, disk radius $9\lambda_0$ and thickness $0.04\lambda_0$. The laser beam waist is $4.5\lambda_0$.}
\label{fig:R-LDOS-delta}
\end{figure}

Finally, by tuning the laser wavelength around the atomic transition and calculating the one-dimensional LDOS \eqref{eq:LDOS}, the opening of forbidden bands in the one-dimensional lattice is observed in Fig.~\ref{fig:R-LDOS-delta}. In fact, we observe a lowering of the LDOS to values of $\sim 0.1$ over a range of a few $\Gamma$s. This value of the LDOS is in agreement with the measurements of Schilke and collaborators~\cite{Schilke2011}, although in that experiment the band was a dozen $\Gamma$s large due to the larger length of the lattice. Again, only an infinite lattice is able to support a perfect band gap with $D=0$.

In conclusion, the microscopic model \eqref{eqbetajstat} captures well one-dimensional forbidden bands: it describes the finite penetration in the lattice, the reflection of the wave and the vanishing LDOS as the lattice size increases.

\subsection{Transfer matrix formalism\label{sec:TM}}

It is customary to study the photonic properties of
one-dimensional optical lattices using the transfer matrix (TM) formalism
~\cite{Deutsch1995}. One of the foundation hypotheses of
the TM theory is that the atomic cloud is assimilated in to a
dielectric medium. Moreover, each lattice site is assumed to be a
radially unlimited slice called a \textit{slab}. This allows one to reduce the
light scattering to the one-dimensional propagation of a plane wave,
where the coupled evolution of the forward and backward waves is
captured in an iterative $2\times2$ matrix problem.

In this section we establish a formal link between the microscopic model~\eqref{eqbetaj} and the TM approach, providing more details of the derivation presented in~\cite{Samoylova2013}. The TM results will also be used as a benchmark to understand the signatures of finite-size effects and atomic disorder in later sections.

Introducing the normalized detuning $\delta=\Delta_{0}/\Gamma$, (~\eqref{eqbetaj}) can be rewritten as:
\begin{equation}
\label{eq:steadystate}
\frac{\Omega_0}{\Gamma}e^{i\textbf{k}_{0}\textbf{r}_{j}}=(2\delta+i)\beta_{j}+i\sum_{k\ne
j}\frac{\exp(ik_{0}|\textbf{r}_{j}-\textbf{r}_{k}|)}{ik_{0}|\textbf{r}_{j}-\textbf{r}_{k}|}\beta_{k}.
\end{equation}
One can then adopt a fluid description of the atomic cloud, i.e., the coarse-grained field $\beta(\mathbf{r})$ describing the local atomic dipole moment field is introduced. This approach has been particularly useful in studying the superradiant and subradiant radiation modes~\cite{Svidzinsky2010}, non-local effects~\cite{Svidzinsky2012}, scattering from clouds with arbitrary spherical densities~\cite{Piovella2012,Piovella2013} and the coherent nature of this scattering~\cite{Prasad2000,Glauber2000,Prasad2010,Prasad2011}. The coarse-grained equation for the cloud excitation in the steady state reads:
\begin{equation}
 \label{eq:FluidEq}
 \frac{\Omega_0}{\Gamma}e^{i\textbf{k}_{0}\textbf{r}}=(2\delta+i)\overline\beta(\textbf{r})+i\int d\textbf{r}' \rho(\textbf{r}')
 \frac{\exp (ik_{0}|\textbf{r}-\textbf{r}'|)}{ik_{0}|\textbf{r}-\textbf{r}'|}\overline\beta(\textbf{r}'),
\end{equation}
where $\rho(\mathbf{r})$ is the atomic density. In the case of
extended slab planes in the transverse directions with density $\rho(z)$
depending only on the longitudinal coordinate, we can
assume $\overline\beta(\textbf{r})=\overline\beta(z)e^{ik_{0y}y}$ where
$k_0^2=k_{0y}^2+k_{0z}^2$, so that (\ref{eq:FluidEq}) can be written as:
\begin{eqnarray}
\label{eq:1Dkernel}
 \frac{\Omega_0}{\Gamma}e^{ik_{0z}z}&=&(2\delta+i)\overline\beta(z)+
 \int dz'\rho(z')\overline\beta(z') \nonumber \\
 &\times &\int dx'\int dy'\frac{\exp (ik_{0}|\textbf{r}-\textbf{r}'|)}
 {k_0|\textbf{r}-\textbf{r}'|}e^{ik_{0y}(y'-y)}.
\end{eqnarray}

By introducing $s=\sqrt{(x'-x)^{2}-(y'-y)^{2}}$ and $s\sin
\phi=y'-y$, the integral over transverse coordinates $x'$ and $y'$
becomes~\cite{Manassah2012}:
\begin{eqnarray}
\int dx'\int
dy'\frac{\exp(ik_{0}|\textbf{r}-\textbf{r}'|)}{|\textbf{r}-\textbf{r}'|}e^{ik_{0y}(y'-y)}&=&
\int\limits_{0}^{\infty}\dfrac{ds
s}{\sqrt{s^2+(z-z')^2}}e^{ik_0\sqrt{s^2+(z-z')^2}}\nonumber
\\ &\times &\int\limits_{0}^{2\pi}d
\phi e^{ik_{0y}s \sin \phi}.
\end{eqnarray}
The integration over $\phi$ and $s$ gives:
\begin{eqnarray}
\int dx'\int dy'\frac{\exp(ik_{0}|\textbf{r}-\textbf{r}'|)}{|\textbf{r}-\textbf{r}'|}e^{ik_{0y}(y'-y)}&=&
2\pi\int\limits_{0}^{\infty}\dfrac{ds s
J_0(k_{0y}s)}{\sqrt{s^2+(z-z')^2}}e^{ik_0\sqrt{s^2+(z-z')^2}} \nonumber
\\ &=&\frac{2\pi i}{k_{0z}}e^{ik_{0z}|z-z'|}.\label{Intxy}
\end{eqnarray}
From Eqs.(\ref{eq:1Dkernel}) and (\ref{Intxy}), the equation for
the 1D scattering problem is:
\begin{equation}
\label{eq:1Dkernelsingleslab}
\frac{\Omega_0}{\Gamma}e^{ik_{0z}z}=(2\delta+i)\overline\beta(z)+\frac{2\pi
i}{k_{0z}k_0}\int dz'\rho(z')\overline\beta(z')e^{ik_{0z}|z-z'|}.
\end{equation}
Then, for an optical lattice consisting of $N_d$ parallel slabs of
uniform density $\rho_0$ and having thickness $a$, separated by a free
space drift $d$ and bounded by the planes $z=0$ and $z=N_da+(N_d-1)d$,
(\ref{eq:1Dkernelsingleslab}) reduces to $N_d$ equations, one for each
slab:
\begin{equation}
 \label{eq:MultiSlices}
  \frac{\Omega_0}{\Gamma}e^{ik_{0z}z}=(2\delta+i)\overline\beta_n(z)+\frac{2\pi i \rho_0}{k_{0z}k_{0}}
  \sum_{m=1}^N \int\limits_{z_{m}}^{z_{m}+a} dz^{'}e^{ik_{0z}|z-z^{'}|}\overline\beta_m(z^{'}),
 \end{equation}
for $z_{n}<z<z_{n}+a$ ($n=1,...N$), where $z_{n}=(n-1)(a+d)$ and
$z_{n}+a$ are the slab edges, while $\overline\beta_n(z)$ refers to the value of
$\overline\beta(z)$ inside the slab. In (\ref{eq:MultiSlices})
$k_{0z}=\sqrt{k_0^{2}-k_{0y}^{2}}=k_0\cos\theta_0$  with
$\theta_0$ being the angle of incidence of the laser with
respect to the lattice axis $z$.

While the one-dimensional problem is usually solved for the
electric field by calculating its propagation layer by layer
(see, e.g., Ref.~\cite{Deutsch1995}), (~\eqref{eq:MultiSlices})
explicitly shows the dependence of the field on the radiation of
each dipole in the system. The collective nature of the scattering
process is thus more evident in our formalism and, although the
derivation is less straightforward, it is formally equivalent to
the standard approach.

Let us now use the fact that the kernel of
(\ref{eq:MultiSlices}) is the Green function for the 1D Helmholtz
equation
\begin{equation}
\left(\partial_{z}^{2}+k_{0z}^{2}\right)e^{ik_{0z}|z-z'|}=2ik_{0z}\delta(z-z').
\end{equation}
Hence, applying the operator $(\partial_{z}^{2}+k_{0z}^{2})$ to (\eqref{eq:MultiSlices}) we obtain the Helmholtz equation which takes the following form inside the $n$th slab:
\begin{equation}
 \label{eq:Helmholtz equation}
 \frac{\partial^{2}\overline\beta_n(z)}{\partial z^{2}}+k_{z}^{2}\overline\beta_n(z)=0,
 \end{equation}
where $k_{z}^{2}=k_{0}^{2}(m_0^{2}-\sin^{2}\theta_{0})$ and $m_0$
is the refractive index of the atomic layer, which can be
expressed as
\begin{equation}\label{eq:m0}
m_0^2=1-\dfrac{4\pi \rho_{0}}{k_{0}^{3}(2\delta+i)}.
\end{equation}
Notice that, according to the Snell-Descartes law ($\sin
\theta_0=m_{0}\sin \theta$), the $z$ component of the wave vector
\textit{inside} the medium can be written as $k_{z}=k_{0}m_{0}\cos
\theta$, where $\theta$ is the angle of refraction. For a
homogeneous slab of refractive index $m_0$ only two modes are
allowed (forward and backward), so the general solution of
(\eqref{eq:Helmholtz equation}) for the $n$th slice is given by
\begin{equation}\label{eq:b(z)}
\overline\beta_n(z)=\dfrac{1}{2\delta+i}\left[
x_{n}e^{ik_{z}(z-z_{n})}+y_{n}e^{-ik_{z}(z-z_{n})}\right].
\end{equation}
The coefficients $x_{n}$ and $y_{n}$ are obtained by substituting
(\ref{eq:b(z)}) into (\ref{eq:MultiSlices}) evaluated at
the boundary conditions $z=z_{n}$ and $z=z_{n}+a$. The long but
straightforward calculation is detailed in Appendix
(\ref{Appendix:xn-yn}).

The recurrence relation for $x_n$ and $y_n$ is given by~\cite{Boedecker2003}
\begin{equation}
\label{T1}
\left[%
\begin{array}{c}
  x_{n+1} \\
  y_{n+1} \\
\end{array}%
\right]= T
\left[%
\begin{array}{c}
  x_{n} \\
  y_{n} \\
\end{array}%
\right],
\end{equation}

where
\begin{equation}
\label{T2}
T=\left[%
\begin{array}{cc}
   T_{11} & T_{12}\\
   T_{21} & T_{22}
   \\
\end{array}%
\right]
\end{equation}
is the transfer matrix with the elements
\begin{eqnarray}
 T_{11,22}&=&\left[\cos(k_{0z}d)\pm i\frac{k_{0z}^{2}+k_{z}^2}{2k_{0z}k_z}\sin(k_{0z}d)\right] e^{\pm ik_z a},\nonumber \\
 T_{12,21}&=&\pm i\frac{k_{0z}^{2}-k_{z}^2}{2k_{0z}k_z}\sin(k_{0z}d)e^{\mp ik_z a}.\label{eq:coeffT}
\end{eqnarray}
Since $\det(T)=1$, its eigenvalues can be written in the form $\lambda_{\pm}=e^{\pm i\phi}$, where $\cos \phi=\textrm{Tr}(T)/2$, or explicitly
\begin{equation}
\cos\phi=\left[\cos(k_{0z}d)\cos(k_z a)-\frac{k_{0z}^{2}+k_{z}^2}{2k_{0z}k_z}\sin(k_{0z}d)\sin(k_za)\right].
\end{equation}
In the case of real refractive index $m_{0}$ and if
$|\cos\phi|<1$, the eigenvalues $\lambda_{\pm}$ are on the unit
circle (real $\phi$) and correspond to propagating (extended)
Bloch modes, whereas if $|\cos\phi|>1$, the eigenvalues are real
(imaginary $\phi$) and can be found in band gaps. In atomic
clouds, the index $m_0$ is always complex and so are the
eigenvalues. It is then necessary to resort to the Local Density
of States to properly characterize the band gaps (see later
discussion).

The scattered field in the smooth density limit is readily obtained from (\eqref{Es:3bis}) and reads~\cite{Bachelard2011}:
\begin{equation}\label{eq:EsTM}
E_S=-\frac{d k_{0}^{2}}{4\pi \varepsilon_{0}}\int
d\textbf{r}'\rho(\textbf{r}')\overline\beta(\textbf{r}')\frac{\exp(ik_{0}|\textbf{r}-\textbf{r}'|)}{|\textbf{r}-\textbf{r}'|}.
\end{equation}
Using \eqref{Intxy}, we obtain the following expression for the total electric field:
\begin{eqnarray}\label{eq:EsN}
E(z)&=&\frac{E_{0}}{2}\left[
e^{ik_{0z}z}+i\frac{k_{z}^{2}-k_{0z}^{2}}{2k_{0z}}(2\delta+i) \right. \nonumber \\
&\times &  \left. \sum_{n=1}^N\int \limits_{z_{n}}^{z_{n}+a} dz'e^{ik_{0z}|z-z'|}\overline\beta_{n}(z')\right],
\end{eqnarray}
where $E_{0}$ is the incident laser field amplitude.
(\ref{eq:EsN}) allows extraction of the reflection and
transmission coefficients $r_{N}$ and $t_{N}$, respectively:
\begin{eqnarray}
\label{rN}
r_N&=&\sum_{n=1}^N e^{ik_{0z}z_n}\left\{\frac{k_{z}-k_{0z}}{2k_{0z}}\left[e^{i(k_{0z}+k_{z})a}-1\right]x_n\right. \nonumber \\
 &-&\left.\frac{k_{z}+k_{0z}}{2k_{0z}}\left[e^{i(k_{0z}-k_{z})a}-1\right] y_n\right\},
\end{eqnarray}
\begin{eqnarray}
\label{tN}
 t_N=1&+&\sum_{n=1}^N e^{-ik_{0z}z_n}\left\{\frac{k_{z}+k_{0z}}{2k_{0z}}\left[e^{i(k_{z}-k_{0z})a}-1\right] x_n\right.\nonumber \\
 &-&\left.\frac{k_{z}-k_{0z}}{2k_{0z}}\left[e^{-i(k_{z}+k_{0z})a}-1\right]y_n\right\}.
\end{eqnarray}

Iterating (\ref{T1}) we can write
\begin{equation}\label{Tn}
\left[%
\begin{array}{c}
  x_{n+1} \\
  y_{n+1} \\
\end{array}%
\right]= T^n
\left[%
\begin{array}{c}
  x_{1} \\
  y_{1} \\
\end{array}%
\right].
\end{equation}
The matrix $T^{n}$ can be easily calculated in the Bloch basis \cite{Boedecker2003}:
\begin{equation}
T^n=M\left[
\begin{array}{cc}
   e^{in\phi}& 0\\
   0 & e^{-in\phi}\\
\end{array}
    \right]M^{-1},
\end{equation}
with
\begin{equation}
M=
\left[%
\begin{array}{cc}
  N_+ & N_-\\
  N_+c_+ & N_-c_{-}\\
\end{array}%
\right],
\end{equation}
where $c_{\pm}=(\lambda_{\pm}-T_{11})/T_{12}$ and $N_{\pm}$ are normalization constants chosen for the Bloch eigenstates.
The explicitly written elements of the matrix $T^{n}$ have the following form:
\begin{eqnarray}
T_{11,22}^{n}&=&\frac{1}{\sin\phi}\left[T_{11,22}\sin(n\phi)-\sin[(n-1)\phi\right], \nonumber \\
T_{12,21}^{n}&=&\frac{1}{\sin\phi}T_{12,21}\sin(n\phi).
\end{eqnarray}

The values $x_1$ and $y_1$ appearing in (\ref{Tn}) can be
determined explicitly from (\ref{eq:MultiSlices}) by
considering the first slab $n=1$ and the last slab $n=N$ and
(\ref{Tn}) at $n=N-1$ (see the Appendix), which allows
rewriting the expressions (\ref{rN}) and (\ref{tN}) for the
reflection and transmission coefficients $r_N$ and $t_N$ in terms
of the reflection and transmission coefficient amplitudes $r$ and
$t$ for a single slab:
\begin{eqnarray}
t_N&=&\frac{t\sin\phi}{\sin N\phi-t\sin(N-1)\phi},\label{eq:tN}\\
r_N&=&t_N\frac{r}{t}\frac{\sin N\phi}{\sin\phi},\label{eq:rN}
\end{eqnarray}
where
\begin{eqnarray}\label{Eq:r}
r&=&\frac{k_{0z}-k_z}{k_{0z}+k_z}\frac{1-e^{2ik_za}}{1-\left(\frac{k_{0z}-k_z}{k_{0z}+k_z}\right )^{2}e^{2ik_za}}\\
t&=&e^{i(k_{0z}d+k_za)}\frac{1-\left(\frac{k_{0z}-k_z}{k_{0z}+k_z}\right)^{2}}{1-\left(\frac{k_{0z}-k_z}{k_{0z}+k_z}\right )^{2}e^{2ik_za}}.\label{Eq:t}
\end{eqnarray}

\subsection{Vectorial microscopic theory\label{sec:vectorial}}

\subsubsection{Discrete model}
Up to now we have considered a scalar cooperative scattering
theory, where the vectorial nature of the electromagnetic field is
disregarded. The scalar theory remains valid for s-polarized light
incident on a 1D periodic stack of planes and is partially
satisfactory for disordered systems, where polarization effects
have a minor role due to the random orientation of the atomic
dipoles. However, the scalar theory does not accurately describe
scattering by periodic systems of higher dimensions, where
a vectorial description is required. The vectorial model of light scattering
can be derived from a quantum field theory (see for instance
Ref.\cite{Manassah2012}) considering electric dipole transitions
between a singlet ground state and a degenerate triplet excited
state (as for instance in a $J=0\rightarrow J=1$ transition).
Identifying $z$ as a quantization axis, the equations for the
three components of the complex polarization vector of the
$j$th atom are given by (see Appendix \ref{Appendix:vectorial}):
\begin{equation}\label{betajk}
    \dot\beta_j^{(\alpha)}=\left(i\Delta_0-\frac{\Gamma}{3}\right)\beta_j^{(\alpha)}
   -i\frac{dE_{0}}{2\hbar}\hat e_{0\alpha} e^{i\mathbf{k}_0\cdot \mathbf{r}_j}
    -\frac{\Gamma}{2}\sum_{\beta}\sum_{m\neq j}G_{\alpha,\beta}(\mathbf{r}_{jm})\beta_m^{(\beta)},
\end{equation}
where $\alpha=(x,y,z)$,
$\mathbf{r}_{jm}=\mathbf{r}_j-\mathbf{r}_m$,
\begin{eqnarray}\label{Gjm}
G_{\alpha,\beta}(\mathbf{r})&=&
    \frac{e^{ik_0r}}{ik_0r}
  \left\{\left[\delta_{\alpha,\beta}-\hat r_\alpha\hat
  r_\beta\right]
  +\left[\delta_{\alpha,\beta}-3\hat
r_\alpha\hat r_\beta\right]\left[\frac{i}{k_0r}-\frac{1}{(k_0r)^2}\right]\right\}\nonumber\\
\end{eqnarray}
and $\mathbf{\hat r}=\mathbf{r}/r$. The incident laser beam has
electric field $\mathbf{E}_{in}(\mathbf{r},t)=E_0\mathbf{\hat
e}_0\cos(i\mathbf{k}_0\cdot \mathbf{r}-i\omega_0t)$ with
polarization unit vector $\mathbf{\hat e}_0$, wave vector
$\mathbf{k}_0$ and frequency $\omega_0=ck_0$ near the atomic
transition frequency $\omega_a$. The vectorial Green function
$G_{\alpha,\beta}(\mathbf{r})$ can be written as a function of the
scalar Green function $G(r)=\exp(ik_0r)/(ik_0r)$ in the following form:
\begin{equation}\label{GvGs}
G_{\alpha,\beta}(\mathbf{r})=\left[\delta_{\alpha,\beta}+\frac{1}{k_0^2}\frac{\partial^2}{\partial
x_\alpha\partial x_{\beta}}\right]G(r)={\cal D}_{\alpha,\beta}G(r),
\end{equation}
where $x_\alpha=(\mathbf{r})_\alpha$. The steady-state problem
reduces to solving the coupled equations:
\begin{equation}\label{betasta}
   \frac{dE_{0}}{\hbar\Gamma}\hat e_{0\alpha}e^{i\mathbf{k}_0\cdot \mathbf{r}_j}=\left(2\delta+\frac{2}{3}i\right)\beta_j^{(\alpha)}
    +i\sum_{\beta}\sum_{m\neq j}G_{\alpha,\beta}(\mathbf{r}_{jm})\beta_m^{(\beta)}.
\end{equation}
This equation is similar to \eqref{eqbetajstat}. The only major
difference is the $(2/3)$ coefficient appearing in the
self-decay term due to single-atom spontaneous emission
isotropy.

\subsubsection{Continuous model}

Neglecting granularity, (\ref{betasta}) can be converted into a
continuous integral equation for the vectorial field
$\overline{\boldsymbol{\beta}}(\mathbf{r})$ with the components
$(\overline{\boldsymbol{\beta}})_\alpha=\overline\beta_\alpha$:
\begin{eqnarray}\label{betavec2}
    \frac{d\mathbf{E}_{0}}{\hbar\Gamma}e^{i\mathbf{k}_0\cdot
    \mathbf{r}}&=&\left(2\delta+2i/3\right)\overline{\boldsymbol{\beta}}\nonumber\\
    &+&i\int d\mathbf{r}'\rho(\mathbf{r}')
    \left\{
    1+\frac{1}{k_0^2}
    \mathbf{\nabla}_{\mathbf{r}}\mathbf{\nabla}_{\mathbf{r}}
    \right\}G(|\mathbf{r}-\mathbf{r}'|)\overline{\boldsymbol{\beta}}(\mathbf{r}'),
\end{eqnarray}
where $\mathbf{E}_0=E_0\mathbf{\hat e}_0$ and $\rho(\mathbf{r})$
is the atomic density. Since
$\mathbf{\nabla}_{\mathbf{r}}\mathbf{\nabla}_{\mathbf{r}}G(|\mathbf{r}-\mathbf{r}'|)
=\mathbf{\nabla}_{\mathbf{r}'}\mathbf{\nabla}_{\mathbf{r}'}G(|\mathbf{r}-\mathbf{r}'|)$,
by integrating by parts the last term of (\ref{betavec2}) we
obtain:
\begin{eqnarray}\label{betavec31}
    \frac{d\mathbf{E}_{0}}{\hbar\Gamma}e^{i\mathbf{k}_0\cdot
    \mathbf{r}}&=&\left(2\delta+2i/3\right)\overline{\boldsymbol{\beta}}\nonumber\\
    &+& i\int d\mathbf{r}'\rho(\mathbf{r}')
    G(|\mathbf{r}-\mathbf{r}'|)
    \left\{\overline{\boldsymbol{\beta}}(\mathbf{r}')+\frac{1}{k_0^2}
    \mathbf{\nabla}_{\mathbf{r}'}\left[\mathbf{\nabla}_{\mathbf{r}'}\cdot \overline{\boldsymbol{\beta}}(\mathbf{r}')
    \right]\right\}.
\end{eqnarray}
Since
\begin{equation}\label{green}
(\nabla^2+k_0^2)G(|\mathbf{r}-\mathbf{r}'|)=\frac{4\pi
i}{k_0}\delta(\mathbf{r}-\mathbf{r}')
\end{equation}
and $(\nabla^2+k_0^2)\exp(i\mathbf{k}_0\cdot \mathbf{r})=0$,
applying the operator $(\nabla^2+k_0^2)$ to both sides of
(\ref{betavec31}), we obtain the following differential equation:
\begin{equation}\label{betavecA}
    \left[\nabla^2+k_0^2m^2(\mathbf{r})\right]\overline{\boldsymbol{\beta}}(\mathbf{r})
    =\left[1-m^2(\mathbf{r})\right]
    \nabla\left[\nabla\cdot \overline{\boldsymbol{\beta}}(\mathbf{r})
    \right],
\end{equation}
where $m(\mathbf{r})$ is the cloud refractive index given by
\begin{equation}\label{m2}
    m^2(\mathbf{r})=1-\frac{2\pi\rho(\mathbf{r})}{k_0^3(\delta+i/3)}.
\end{equation}
We notice that in the scalar radiation theory (\ref{betavecA})
reduces to
$[\nabla^2+k_0^2m^2(\mathbf{r})]\overline\beta_\alpha(\mathbf{r})=0$
for each $\alpha$-components. Hence, in the scalar approximation
$\overline{\boldsymbol{\beta}}$ is a purely transverse field that satisfies
$\nabla\cdot \overline{\boldsymbol{\beta}}=0$.

\subsubsection{Scattered field}

The electric field of the scattered radiation can be evaluated
directly from the macroscopic Maxwell equations with polarization
$\mathbf{P}=-d\sum_{j=1}^N{\boldsymbol{\beta}}_j\delta(\mathbf{r}-\mathbf{r}_j)$.
The result, as demonstrated in Appendix \ref{Appendix:field},
is
\begin{equation}\label{Evetto}
E_\alpha(\mathbf{r})=-i\frac{dk_0^3}{4\pi\epsilon_0}\sum_{\beta}\sum_{j=1}^N
G_{\alpha,\beta}(\mathbf{r}-\mathbf{r}_j) \beta^{(\beta)}_j.
\end{equation}
Considering $G_{\alpha,\beta}$ as a spatial component of the
symmetric tensor $\mathbf{G}$, (\ref{Evetto}) can be written in
the vectorial form:
\begin{equation}\label{Evec}
\mathbf{E}(\mathbf{r})=-i\frac{dk_0^3}{4\pi\epsilon_0}\sum_{j=1}^N
\mathbf{G}(\mathbf{r}-\mathbf{r}_{j})\cdot{\boldsymbol{\beta}}_j.
\end{equation}
 Combining (\ref{Evetto}) and (\ref{betasta})
we find:
\begin{equation}\label{beta+E}
   {\boldsymbol{\beta}}_j=\frac{d}{\hbar\left(\Delta_0+i\Gamma/3\right)}
   \left[\mathbf{E}_{in}(\mathbf{r}_j)+\mathbf{E}(\mathbf{r}_j)\right],
\end{equation}
where
$\mathbf{E}_{in}(\mathbf{r})=(\mathbf{E}_0/2)e^{i\mathbf{k}_0\cdot
\mathbf{r}}$ is the incident field and $j=1,\dots,N$. As expected,
the electric dipole moment of the single atom is proportional to
the sum of the incident field and the field scattered by all the
other atoms.

\subsubsection{Single-scattering contribution}

Neglecting the contribution due to the internal field
$\mathbf{E}(\mathbf{r}_j)$ in (\ref{beta+E}) (so, the effect of multiple scattering is not taken into account) from
(\ref{Evec}) and (\ref{beta+E}) we obtain:
\begin{equation}\label{E1}
    \mathbf{E}^{(1)}(\mathbf{r})=\kappa(\delta)\sum_j
    \mathbf{G}(\mathbf{r}-\mathbf{\mathbf{r}}_{j})\cdot\mathbf{E}_{in}(\mathbf{r}_j),
\end{equation}
where $\kappa(\delta)=1/(2i\delta-2/3)$. Far away from the
scatters we can approximate $|\mathbf{r}-\mathbf{r}_{j}|\approx
r-\mathbf{\hat n}\cdot \mathbf{r}_j$ in the exponent factor of
(\ref{Gjm}), where $\mathbf{\hat n}$  is a unit vector in the
direction of the observation and $r$ is the distance to the
system of scatters. Retaining only the terms decreasing as
$1/|\mathbf{r}-\mathbf{r}_{j}|\approx 1/r$, (\ref{Gjm}) is
approximated by
\begin{equation}\label{Gfar}
    G_{\alpha,\beta}(\mathbf{r}-\mathbf{r}_j)\approx
    \frac{e^{ik_0r}}{ik_0r}
  \left(\delta_{\alpha,\beta}-\hat n_\alpha\hat n_\beta\right)e^{-i\mathbf{k}\cdot \mathbf{r}_j},
\end{equation}
where $\mathbf{k}=k_0\mathbf{\hat n}$. Inserting (\ref{Gfar})
in (\ref{E1}) we obtain:
\begin{equation}\label{E1far}
    \mathbf{E}^{(1)}(\mathbf{r})\approx i\kappa(\delta)\frac{e^{ik_0r}}{2k_0r}E_0
    [\mathbf{\hat n}\times(\mathbf{\hat n}\times\mathbf{\hat e}_{0})]
    \sum_{j=1}^N e^{i(\mathbf{k}_0-\mathbf{k})\cdot \mathbf{r}_j},
\end{equation}
where $\mathbf{\hat e}_{0}$ is the incident polarization vector.
Equation (\ref{E1far}) coincides with the well-known expression for the
radiation field emitted by a collection of point-like scatters,
known as Rayleigh scattering~\cite{Jackson1999}. The scattered
field results from a coherent superposition of the field
amplitudes generated by each atom and is proportional to the
structure factor ${\cal F}(\mathbf{q})=\sum_j\exp(i\mathbf{q}\cdot
\mathbf{r}_j)$, where $\mathbf{q}=\mathbf{k}_0-\mathbf{k}$ is the
transferred momentum. Equation (\ref{E1far}) also shows that the
electric dipole moment of each atom is
$\mathbf{p}=\alpha\epsilon_0\mathbf{E}_{in}$ with polarizability
$\alpha(\delta)=4\pi i\kappa(\delta)/k_0^3$.

\subsection{Photonic band gaps for three-level atoms\label{sec:3level}}

The microscopic model of cooperative scattering and photonic band
gaps, so far developed only for two-level atoms, can be extended to
three-level atoms in $\Lambda$ or cascade configurations. These
systems offer the advantage of further manipulations and control
of the photonic band gaps, exploiting the electromagnetically
induced transparency (EIT) properties~\cite{Harris1997}. EIT is a quantum interference effect
characterized by the presence of a frequency region where
absorption is greatly reduced, accompanied by steep dispersion~\cite{Fleischhauer2005}. Both the transparency bandwidth and the steep
dispersion near the EIT resonance are controlled by the
corresponding driving field~\cite{Petrosyan2007}. The possible advantages and limitations of using
the photonic band gaps near the EIT transparency bandwidth has
been investigated in a recent experiment~\cite{Schilke2012a}. Here we develop a microscopic description
of the scattering by three-level atoms, showing how the TM formalism can take into account EIT, simply by
modifying the atomic medium refractive index.

\subsubsection{Single-particle dynamics}

\begin{figure}
        \centerline{\scalebox{0.7}{\includegraphics{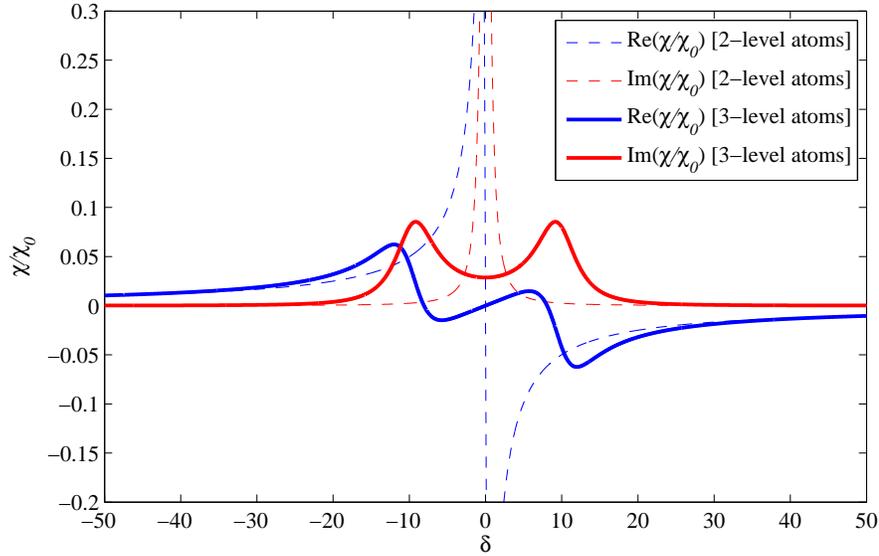}}}
        \caption{Real (blue) and imaginary (red) parts of the normalized susceptibility for two- (dashed) and three-level (solid) atoms. For three-level atoms the following parameters are used: $a=20$ and $\gamma=11.8$.}
      \label{fig1321}
    \end{figure}

We consider a cascade configuration between the three states
$|g\rangle$, $|e\rangle$ and $|m\rangle$ with energy differences
$\hbar\omega_{eg}=E_e-E_g$ and $\hbar\omega_{me}=E_m-E_e$ and
decay rates $\Gamma_{eg}$ and $\Gamma_{me}$,  respectively. For
instance, for $^{88}$Sr atoms the three states may be
$|g\rangle=(5s^2)^1S_0$, $|e\rangle=(5s5p)^3P_1$ and
$|m\rangle=(5s5d)^3D_1$. The transition
$|m\rangle\rightarrow|g\rangle$ is forbidden and the two
transitions $|g\rangle\rightarrow|e\rangle$ and
$|e\rangle\rightarrow|m\rangle$ are driven by two external fields
with Rabi frequencies $\Omega_{1}=d_{ge}E_1/\hbar$,
$\Omega_{2}=d_{em}E_2/\hbar$ and frequencies $\omega_{1,2}$, where $d_{ab}$ is the dipole matrix elements for the
generic transition $|a\rangle\rightarrow |b\rangle$. The
single-particle Hamiltonian is
\begin{eqnarray}\label{H1}
H&=&\hbar\omega_{eg}|e\rangle\langle
e|+\hbar(\omega_{eg}+\omega_{me})|m\rangle\langle
m|\nonumber\\
&-&\hbar\left\{\Omega_1e^{-i\omega_1t}|e\rangle\langle
g|+\Omega_2e^{-i\omega_2t}|m\rangle\langle
e|+\textrm{h.c.}\right\}.
\end{eqnarray}
Assuming
$|\psi\rangle=c_g(t)|g\rangle+c_e(t)|e\rangle+c_m(t)|m\rangle$,
from the Schr\"{o}dinger equation,
$i\hbar\partial_t|\psi\rangle=H|\psi\rangle$, we obtain:
\begin{eqnarray}
  \dot c_g &=&  i\Omega_1^*e^{i\omega_1 t}c_e,\\
  \dot c_e &=&  -i\omega_{eg}c_e+i\Omega_1e^{-i\omega_1 t}c_g+i\Omega_2^*e^{i\omega_2 t}c_m,\\
  \dot c_m &=& -i(\omega_{eg}+\omega_{me})c_m+i\Omega_2e^{-i\omega_2 t}c_e.
\end{eqnarray}
Introducing the coherences $\rho_{ge}=c_g^*c_e \exp(i\omega_1t)$,
$\rho_{gm}=c_g^*c_m\exp[i(\omega_1+\omega_2)t]$ and
$\rho_{em}=c_e^*c_m\exp(i\omega_2t)$, we write:
\begin{eqnarray}
  \dot \rho_{ge} &=&  [i\Delta_1-\Gamma_{eg}/2]\rho_{ge}+i\Omega_1(\rho_{gg}-\rho_{ee})+i\Omega_2^*\rho_{gm}\label{rhoge},\\
  \dot \rho_{gm} &=&  [i(\Delta_1+\Delta_2)-\Gamma_{me}/2]\rho_{gm}-i\Omega_1\rho_{em}+i\Omega_2\rho_{ge}\label{rhogm},\\
  \dot \rho_{em} &=& [i\Delta_2-(\Gamma_{eg}+\Gamma_{me})/2]\rho_{em}+i\Omega_2(\rho_{ee}-\rho_{mm})-i\Omega_1^*\rho_{gm}\label{rhoem},
\end{eqnarray}
where $\rho_{ii}=|c_i|^2$, $\Delta_1=\omega_1-\omega_{eg}$ and
$\Delta_2=\omega_2-\omega_{me}$, and we add the decay terms
with rates $\Gamma_{eg}$ and  $\Gamma_{me}$. We assume the field
$\Omega_1$ to be so weak that $\rho_{gg}\approx 1$, $\rho_{ee}\approx
0$ and $\rho_{mm}\approx 0$. With these approximations
(\ref{rhoem}) yields at steady-state
\[
\rho_{em}\approx\frac{\Omega_1^*\rho_{gm}}{\Delta_2+i(\Gamma_{eg}+\Gamma_{me})/2},
\]
so that $\rho_{em}$ can be neglected in (\ref{rhogm}). The latter then yields
\[
\rho_{gm}\approx-\frac{\Omega_2\rho_{ge}}{(\Delta_1+\Delta_2)+i\Gamma_{me}/2}
\]
and can be inserted into \eqref{rhoge} to obtain
\begin{equation}\label{rhoge:1}
\rho_{ge}
=-\frac{\Omega_1}{\left[\Delta_1+i\Gamma_{eg}/2-\frac{|\Omega_2|^2}{(\Delta_1+\Delta_2)+i\Gamma_{me}/2}\right]}.
\end{equation}
Defining the polarization $P=\rho d_{ge}\rho_{ge}=\epsilon_0\chi
E_1$, where $\rho$ is the atomic density, we obtain the following expression:
\begin{equation}\label{chi}
    \chi=-\frac{\chi_0}{\left[2\delta+i
    -\frac{a^2}{2\delta+i\gamma}\right]},
\end{equation}
where $\chi_0=(2nd_{ge}^2)/(\epsilon_0\hbar\Gamma_{eg})$,
$\delta=\Delta_1/\Gamma_{eg}$, $\gamma=\Gamma_{me}/\Gamma_{eg}$,
$a=2\Omega_2/\Gamma_{eg}$ and we assumed $\Delta_2=0$. Separating
the real and imaginary parts, we write:
\begin{eqnarray}
  \textrm{Re}\left(\frac{\chi}{\chi_0}\right) &=& -2\delta\frac{4\delta^2-\gamma^2-a^2}{(4\delta^2-\gamma-a^2)^2+4\delta^2(1+\gamma)^2}, \label{rechi}\\
  \textrm{Im}\left(\frac{\chi}{\chi_0}\right) &=&
  \frac{\gamma(\gamma+a^2)+4\delta^2}{(4\delta^2-\gamma-a^2)^2+4\delta^2(1+\gamma)^2}\label{imchi}.
\end{eqnarray}
For instance, for $^{88}$Sr,  $\Gamma_{eg}=(2\pi)7.6$kHz and
$\Gamma_{me}=(2\pi)90.3$kHz, so that $\gamma=11.8$. In Fig.~\ref{fig1321} the real and imaginary parts of $\chi/\chi_0$ are represented
as a function of $\delta$ for two- and three-level atoms.
For three-level atoms the imaginary part of $\chi$ has a minimum near
$\delta=0$ provided $a^2>\gamma^3/(1+2\gamma)$, which for
$\gamma=11.8$ gives $\Omega_2>4.1\Gamma_{eg}$.

\subsubsection{Cooperative emission}

\begin{figure}
\centerline{\scalebox{0.7}{\includegraphics{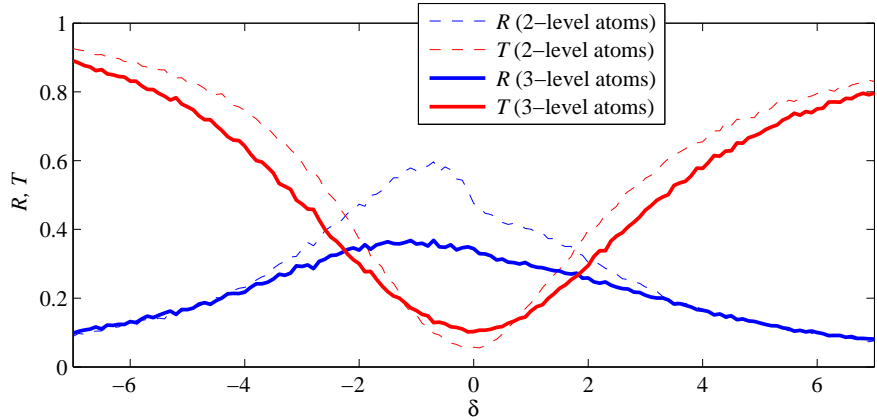}}}
\caption{Reflection $R$ and transmission $T$ coefficients for two- and three-level atom lattices. The lattice is composed of $8000$ atoms spread over $200$ disks of thickness $0.04\lambda_0$ and radius $4.5\lambda_0$. For the three-level atoms we choose $a=4$ and $\gamma=11.8$.}
      \label{fig:RT3lvl}
    \end{figure}

Once the single-particle dynamics have been determined it is
straightforward to obtain the cooperative scattering model for the
three-level cascade configuration. Considering a collection of
such $N$ three-level atoms, the Hamiltonian of the system in the
scalar field theory, including the coupling with the vacuum
radiation modes, is given by
\begin{eqnarray}\label{HN}
H&=&\hbar\sum_{j=1}^N\left\{\omega_{eg}|e_j\rangle\langle
e_j|+(\omega_{eg}+\omega_{me})|m_j\rangle\langle
m_j|\right\}\nonumber\\
&-&\hbar\sum_{j=1}^N\left\{\Omega_1e^{-i\omega_1t+i\mathbf{k}_1\cdot
\mathbf{r}_j}|e_j\rangle\langle
g_j|+\Omega_2e^{-i\omega_2t+i\mathbf{k}_2\cdot
\mathbf{r}_j}|m_j\rangle\langle
e_j|+\textrm{h.c.}\right\}\nonumber\\
&+&\hbar\sum_{j=1}^N\sum_{\mathbf{k}}g_{\mathbf{k}}
\{a_{\mathbf{k}} e^{-i\omega_kt+i\mathbf{k}\cdot
\mathbf{r}_j}+a_{\mathbf{k}}^\dagger
e^{i\omega_kt-i\mathbf{k}\cdot \mathbf{r}_j}\}\{|e_j\rangle\langle
g_j|+|g_j\rangle\langle e_j|)\nonumber\\
\end{eqnarray}
We assume that the cooperative effects manifest themselves only in
the lower transition $|e\rangle\rightarrow |g\rangle$, as
described by the last term in the Hamiltonian (\ref{HN}). The
single-atom spontaneous emission from the upper level $|m\rangle$
is described simply by adding a damping term, as performed in the previous section. Assuming at most one atom in the excited state $|e\rangle$, the complete system
is described by the following state:
\begin{eqnarray}\label{psitot}
    |\psi\rangle&=&c_g(t)|g_1,\dots,g_N\rangle|0\rangle_{\mathbf{k}}+\sum_{j=1}^Nc_{ej}(t)|g_1,\dots,e_j,\dots,g_N\rangle|0\rangle_{\mathbf{k}}
    \nonumber\\
    &+&
    \sum_{j=1}^Nc_{mj}(t)|g_1,\dots,m_j,\dots,g_N\rangle|0\rangle_{\mathbf{k}}\nonumber\\
    &+&\sum_{\mathbf{k}}\gamma_\mathbf{k}(t)|g_1,\dots,g_N\rangle|1\rangle_{\mathbf{k}}\nonumber\\
&+&\sum_{\mathbf{k}}\sum_{j\neq
j'}\epsilon_{j,j',\mathbf{k}}(t)|g_1,\dots,e_j,\dots,e_{j'},\dots,g_N\rangle|1\rangle_{\mathbf{k}}.
\end{eqnarray}
The evolution of the probability amplitudes is governed by:
\begin{eqnarray}
  \dot c_{g}(t) &=&  i\Omega_1^*\sum_{j=1}^Nc_{ej}(t)e^{i\omega_1 t-i\mathbf{k}_1\cdot \mathbf{r}_j},\\
  \dot c_{ej}(t) &=&  -i\omega_{eg}c_e(t)+i\Omega_1 c_g(t) e^{-i\omega_1 t+i\mathbf{k}_1\cdot \mathbf{r}_j}
  +i\Omega_2^* c_m(t)e^{i\omega_2
  t-i\mathbf{k}_2\cdot \mathbf{r}_j}\nonumber\\
  &-& i\sum_{\mathbf{k}}g_{\mathbf{k}}\left[\gamma_{\mathbf{k}}(t)+\sum_{j'\neq j}
  \epsilon_{j,j',\mathbf{k}}(t)\right]e^{-i\omega_k t+i\mathbf{k}\cdot \mathbf{r}_j},\\
  \dot c_{mj}(t) &=& -i(\omega_{eg}+\omega_{me})c_{mj}(t)+i\Omega_2 c_{ej}(t) e^{-i\omega_2
  t+i\mathbf{k}_2\cdot \mathbf{r}_j},\\
  \dot\gamma_{\mathbf{k}}(t)&=&-ig_{\mathbf{k}}\sum_{j=1}^Nc_{ej}(t)e^{i\omega_k t-i\mathbf{k}\cdot
  \mathbf{r}_j}\\
  \dot\epsilon_{j,j'\mathbf{k}}(t)&=&-ig_{\mathbf{k}}e^{i\omega_k t}
  \left[e^{-i\mathbf{k}\cdot \mathbf{r}_j}c_{ej'}(t)+e^{-i\mathbf{k}\cdot \mathbf{r}_{j'}}c_{ej}(t)\right].
\end{eqnarray}
Introducing again the coherences and assuming the field $\Omega_1$
to be weak and the ground state undepleted, we obtain:
\begin{eqnarray}
  \dot \rho_{ge}^{(j)}(t) &=&  i\Delta_1\rho_{ge}^{(j)}(t)+i\Omega_1e^{i\mathbf{k}_1\cdot \mathbf{r}_j}
  +i\Omega_2^*e^{-i\mathbf{k}_2\cdot
  \mathbf{r}_j}\rho_{gm}^{(j)}(t)\nonumber\\
  &-&i\sum_{\mathbf{k}}g_{\mathbf{k}}\left[\gamma_{g\mathbf{k}}(t)+\sum_{j'\neq j}\epsilon_{g,\mathbf{k}}^{(j,j')}(t)\right]e^{-i(\omega_k-\omega_1)t+i\mathbf{k}\cdot \mathbf{r}_j},\label{rhogej}\\
  \dot \rho_{gm}^{(j)}(t) &=& [i(\Delta_1+\Delta_2)-\Gamma_{me}/2]\rho_{gm}^{(j)}(t)+i\Omega_2e^{i\mathbf{k}_2\cdot \mathbf{r}_j}\rho_{ge}^{(j)}(t),
  \label{rhogmj}\\
  \dot\gamma_{g\mathbf{k}}(t)&=&-ig_{\mathbf{k}}\sum_{j=1}^N\rho_{ge}^{(j)}(t)e^{i(\omega_k-\omega_1) t-i\mathbf{k}\cdot
  \mathbf{r}_j},\\
  \dot\epsilon_{g\mathbf{k}}^{(j,j')}(t)&=&-ig_{\mathbf{k}}e^{i(\omega_k-\omega_1)t}
  \left[e^{-i\mathbf{k}\cdot
\mathbf{r}_j}\rho_{ge}^{(j')}(t)+e^{-i\mathbf{k}\cdot
\mathbf{r}_{j'}}\rho_{ge}^{(j)}(t)\right],
\end{eqnarray}
where $\gamma_{g\mathbf{k}}=c_g^*\gamma_{\mathbf{k}}$ and
$\epsilon_{g\mathbf{k}}^{(j,j')}=c_g^*\epsilon_{j,j',\mathbf{k}}$.
As we see, the only difference with respect to the two-level atom
case is the third term on the right hand side of (\ref{rhogej})
coupled to the upper transition by the field $\Omega_2$. By
eliminating the variables $\gamma_{g\mathbf{k}}$ and
$\epsilon_{g\mathbf{k}}^{(j,j')}$ in the usual Markov
approximation, (\ref{rhogej}) becomes
\begin{eqnarray}\label{beta3j}
    \dot \rho_{ge}^{(j)}(t) &=&  i\Delta_1\rho_{ge}^{(j)}(t)+i\Omega_1e^{i\mathbf{k}_1\cdot \mathbf{r}_j}
  +i\Omega_2^*e^{-i\mathbf{k}_2\cdot
  \mathbf{r}_j}\rho_{gm}^{(j)}(t)\nonumber\\
  &-&\frac{\Gamma_{eg}}{2}\sum_{m=1}^N\frac{\exp[ik_1|\mathbf{r}_j-\mathbf{r}_m|]}{ik_1|\mathbf{r}_j-\mathbf{r}_m|}\rho_{ge}^{(m)}(t).
\end{eqnarray}
Neglecting granularity and assuming a continuous density
distribution $\rho(\mathbf{r})$,  (\ref{rhogmj}) at
steady-state reads
\[
\rho_{gm}(\mathbf{r})=-\frac{\Omega_2e^{i\mathbf{k}_2\cdot
\mathbf{r}_j}}{(\Delta_1+\Delta_2)+i\Gamma_{me}/2}\rho_{ge}(\mathbf{r}).
\]
Inserted in to \eqref{beta3j}, this gives
\begin{eqnarray}\label{beta3ss}
   -\Omega_1e^{i\mathbf{k}_1\cdot \mathbf{r}}&=&  \left[\Delta_1+i\frac{\Gamma_{eg}}{2}
  -\frac{|\Omega_2|^2}{\Delta_1+\Delta_2+i\Gamma_{me}/2}\right]\rho_{ge}(\mathbf{r})\nonumber\\
  &+&\frac{\Gamma_{eg}}{2}\int d\mathbf{r}'\rho(\mathbf{r}')\frac{\exp(ik_1|\mathbf{r}-\mathbf{r}'|)}{k_1|\mathbf{r}-\mathbf{r}'|}\rho_{ge}(\mathbf{r}').
\end{eqnarray}
Assuming for simplicity $\Delta_2=0$ and defining
$\delta=\Delta_1/\Gamma_{eg}$, $a=2|\Omega_2|/\Gamma_{eg}$,
$\gamma=\Gamma_{me}/\Gamma_{eg}$ and
$\overline{\beta}(\mathbf{r})=-\rho_{ge}(\mathbf{r})$,
(\ref{beta3ss}) becomes
\begin{eqnarray}\label{beta3ss2}
   \frac{\Omega_1}{\Gamma_{eg}}e^{i\mathbf{k}_1\cdot \mathbf{r}}&=&  \left[2\delta+i
  -\frac{a^2}{2\delta+i\gamma}\right]\overline{\beta}(\mathbf{r})\nonumber\\
  &+&\int
  d\mathbf{r}'\rho(\mathbf{r}')\frac{\exp(ik_1|\mathbf{r}-\mathbf{r}'|)}{k_1|\mathbf{r}-\mathbf{r}'|}\overline{\beta}(\mathbf{r}').
\end{eqnarray}
Equation (\ref{beta3ss2}) generalizes the coarse-grained
(\ref{eq:FluidEq}) for the case of three-level atoms. The only
difference is the extra term in the squared parenthesis,
proportional to the driving intensity. Moreover, the TM formalism remains unchanged, only the
refractive index in (\ref{eq:m0}) is replaced by the expression
\begin{equation}\label{eq:m03}
m_0^2=1-\dfrac{4\pi
\rho_{0}}{k_{1}^{3}\left(2\delta+i-\frac{a^2}{2\delta+i\gamma}\right)}.
\end{equation}
In \eqref{eq:m03}, the possibility to manipulate the cloud's refractive index using a third level becomes clear. An illustration is shown in Fig.\ref{fig:RT3lvl}, where the spectrum in reflection and transmission of a lattice are plotted for two- and three-level atoms.

\section{Measurement of a photonic band gap\label{sec:experiment}}

One-dimensional photonic band gaps in optical lattices have been observed in a recent experiment \cite{Schilke2011}. The basic idea of this experiment was to create a pile of pancake-shaped atomic clouds, to irradiate them with a probe laser tuned close to an atomic resonance at an incident angle satisfying the Bragg condition \eqref{eq:BraggAngle}. The reflected light intensity is then monitored while the probe laser frequency is scanned across resonance to look for signatures of the presence of a band gap in the reflection spectrum.

\subsection{Description of the experiment\label{sec:setup}}

Technically, a cloud of rubidium atoms was trapped and cooled in a standard magneto-optical trap and then transferred into an optical dipole trap formed by a retroreflected laser beam with 1.3~W of power focused into a $220~\mu$m waist and red detuned from the $^{87}$Rb D2-line at $\lambda_0=780.24~$nm. Inside the standing wave formed by the laser beam the atoms arranged themselves at the antinodes and formed a lattice adopting the shape of a pile of pancakes aligned along the optical axis. About 7700 pancakes were filled, each one with on average 6500 atoms. At temperatures around $100~\mu$K the pancakes represented Gaussian density distributions with radii of $\sigma_r=60~\mu$m and thicknesses of $\sigma_z=47~$nm. The optical layout of the experiment is shown in Fig.~\ref{fig:ExpScheme}.

Reflection spectra were obtained by irradiating a probe laser (wavevector $k_{pr}=2\pi/\lambda_{pr}$) under an angle of $\theta=2^\circ$ focused down to $35~\mu$m into the optical lattice. The Bragg condition is satisfied when the difference between the incident and reflected wavevectors matches the lattice wave vector, $2nk_0\cos\theta=4\pi/\lambda_{lat}$, where the real part of the average refractive index $n=n(\delta)$ depends on the probe laser detuning from resonance. However, it is experimentally more convenient to tune the Bragg condition varying the wavelength of the lattice laser, instead of varying the angle of incidence. Introducing the lattice laser detuning from the geometrically ideal ($n=1$) Bragg condition, $\Delta\lambda_{lat}\equiv\lambda_{lat}-\lambda_0/\cos\theta$, the Bragg condition can be written as $n(\delta)-1=-\Delta\lambda_{lat}/\lambda_{lat}$.

For a chosen lattice wavelength, the probe laser was scanned across the $F=2\rightarrow F'=3$ transition of the rubidium D2-resonance, while recording the reflected and transmitted light intensity. As explained in Ref.~\cite{Slama2005}, when the atoms are axially strongly localized in the standing light wave, the atoms scattering into the Bragg angle do not receive recoil, hence the reflected light is not Doppler-shifted, but scattered elastically. As a consequence, the atoms are not heated by the probe beam and the reflected light is not Doppler-broadened. At low atomic densities, the spectral linewidth observed in the reflection spectrum was equal to the natural linewidth of $\Gamma/2\pi=6~$MHz. In contrast, at high atomic densities, when the Bragg condition was met a very efficient reflection of up to 80\% was observed as well as a dramatic and asymmetric spectral broadening.

\subsection{Reaching the thick grating regime\label{sec:thickgrating}}

The experimental challenge in reaching the regime where the lattice responds to incident light by the formation of photonic bands arises from the necessity of realizing a \textit{thick grating}, which means that the light is Bragg-reflected back and forth many times between subsequent atomic layers.

This requires, first of all, that the lattice's optical density is high enough for multiple scattering. Second, the atoms need to be well arranged in axial direction, ideally forming two-dimensional sheets exactly located at the antinodes of the optical lattice. Any atom dislocated from this plane reduces specular reflection through diffuse scattering into arbitrary solid angles. Furthermore, the incident light gets absorbed before it has a chance of penetrating deep into the lattice and to probe many atomic layers. The diffuse scattering problem can be partially circumvented by detuning the probe laser slightly from resonance. The main reason for axial disorder is the axial Gaussian distribution of the atoms within a pancake at finite temperatures whose {\it rms}-width is $\sigma_z\propto\sqrt{T}$. One could think of lowering the temperature by some means, but this also reduces the radial extent of the pancakes, $\sigma_r\propto\sqrt{T}$, which introduces another problem.

As discussed in Section \ref{sec:ApplyTMF}, due to the non-zero angle of incidence, the finite radial extent of the pancakes causes the light to walk off the optical axis after a certain number of reflections~\cite{Slama2005a}, which also limits the number of pancakes available to multiple reflections. Choosing a small angle of incidence reduces this problem, but still walk-off losses represent a serious limitation in some circumstances~\cite{Schilke2012}.

\subsection{Interpretation of the observations\label{sec:observations}}

In the low optical density limit (thin grating limit), the light reflection results from Bragg scattering, which is understood as constructive interference of the Rayleigh-scattered radiation patterns emitted by the individual periodically arranged atoms. In this regime the reflection coefficient of the lattice turns out to be nearly real, phase shifts are negligible, and the spectral lineshape, which is determined by the natural decay rate $\Gamma$, is Lorentzian. Consequently, Bragg scattering in this regime is cooperative but not superradiant.

\begin{figure}
    \centerline{\scalebox{0.5}{\includegraphics{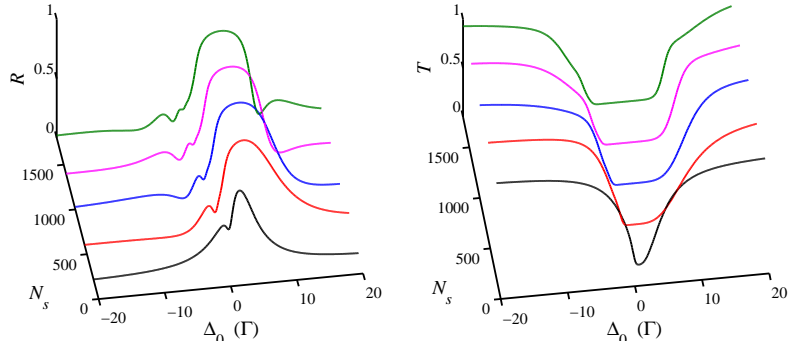}}}
    \caption{Bragg reflection (left) and transmission (right)   spectra  for  lattices  with various  numbers  of  layers $N_d$.  Here  we  assumed  an atomic density $5\times 10^{11}cm^{−3}$. The lattice is slightly detuned from the Bragg condition $\lambda_{dip} = 1 nm$, so that the line shapes become asymmetric.}
    \label{fig:deepMS}
\end{figure}

When the density of the atomic gas, or the optical thickness of the lattice, is increased, the probability that photons are reflected multiple times between adjacent layers also increases (thick grating regime). The interference between the light reflected from or transmitted through the layers gives rise to stopping bands for certain light frequencies or irradiation angles. In this regime absorption can generally be neglected, but large phase shifts occur and the spectral lineshape dramatically broadens under the influence of superradiance and becomes asymmetric (see Fig.~\ref{fig:deepMS}). In the extreme limit of very high densities the broadened reflection spectrum can be interpreted as a frequency band in which the propagation of light traversing the optical lattice under the Bragg angle is forbidden. This frequency band is known as ``photonic band gap''.

To support this interpretation of the observations, Schilke \textit{et al.} compared the observed reflection spectra to the prediction of the TM formalism detailed in Sec.~\ref{sec:TM} and found excellent agreement for all density regimes. The model also allows calculation of the LDOS. For the set of parameters which best fit the reflection spectra a considerable reduction of the LDOS below the free space value of 1 was found.

\section{Finite size effects\label{sec:FiniteSize}}

The TM formalism is broadly used to describe photonic band gaps, yet it may present severe limitations in the case of cold atoms in relatively small optical lattices (as compared to dielectric photonic crystals). Typical three-dimensional lattices may only contain $10^5-10^6$ atoms, which corresponds to several dozens of periods in each direction only. Hence, finite-size effects which can not be accounted for by the TM approach are expected to come into play. Furthermore, while some of the experimental peculiarities and imperfections can be incorporated into a \emph{generalized transfer matrix approach} \cite{Slama2006} other effects are intrinsically beyond it.

\subsection{Applicability of the TM formalism to experiment\label{sec:ApplyTMF}}

Let us now discuss several limitations of the TM formalism with respect to the experimental situation:

1.~If a probe beam enters an optical lattice under a finite angle the beam walks out of the stack of slabs after a finite amount of reflections due to the finite radial distribution. This is similar to the well-known situation in unstable optical cavities. This effect limits the effective number of available slabs contributing to multiple scattering. The effect is particularly pronounced for large angles of incidence or small radial extension of the slabs. While the finite angle of incidence can be incorporated into the formalism via a modified quasi-momentum, the beam walk-off can not because the TM formalism, being purely one-dimensional, assumes not only radially infinite atomic layers, but also a radially homogeneous density distribution. In reality the radial density distribution is rather Gaussian which implying a variation of the penetration depth with the distance from the optical axis. The experimentally observed reflection spectra thus represent an average of reflection spectra taken at {\it different} optical densities.

2.~The finite radial extent has another important impact on the reflection angle. Although the pancake's aspect ratio is smaller than $\sigma_z/\sigma_r\approx 10^{-3}$ with respect to Bragg scattering, it tends to behave like a chain of point-like scatterers rather than a dielectric mirror \cite{Slama2005a}. This means that if the lattice constant is detuned from the Bragg condition, the reflection angle tends to self-adjust in order to fulfil the Bragg condition rather than to be equal to the angle of incidence as assumed in the TM formalism. This self-adjustment of the Bragg condition impedes a controlled tuning of quasi-momentum.

3.~Atoms distributed over the lattice potential, the depth of which is on the order of $U_0=h\cdot7~$MHz, experience individual dynamical Stark shifts of their resonances which vary with the atomic location. This effect shifts and inhomogeneously broadens the Bragg spectra \cite{Slama2005}. The axial modulation of the Stark shift can be included into the TM formalism by dividing every atomic layer into a number of thin sublayers for which the transfer matrices are evaluated based on the local density and detuning of the probe light. However, the radial variation of the Stark shift can not be described.

4.~When the probe laser is detuned from the resonance, the average refractive index of the cloud differs from 1. This means that the incident light is deflected by the refraction when it enters the optically thick atomic cloud. This slightly modifies the angle of incidence, and thus its deviation from the Bragg angle can be as much as $0.1^\circ$ depending on the probe beam detuning~\cite{Slama2006}. Moreover, the optically thick cloud focuses or defocuses the incident beam depending on its detuning.

5.~A good collimation of the probe beam is important. A divergent probe beam may be expanded into several plane waves, and each of them might have a slightly different angle of incidence with respect to the Bragg condition. On the other hand, the probe beam needs to have a waist smaller than the radial extent of the atomic cloud in order to yield high reflection coefficients.

6.~Collective effects arising from the disordered part of the atomic cloud (forward scattering, Mie scattering, superradiance, etc.) can play a role even at moderate atomic densities. However, local disorder is disregarded in the TM formalism. Only disorder along the optical axis of the 1D lattice (finite Debye-Waller factor) could be introduced into the TM formalism.

7.~As already mentioned, the atoms are axially strongly localized by the lattice potential, so that the Bragg-reflected light is scattered elastically due to the Lamb-Dicke effect. This does not hold for diffuse scattering into other solid angles, which is inelastic as it involves the transfer of photonic recoil to atoms in radial direction. The approach of dividing the atomic cloud into a perfectly ordered optical lattice and a separate homogeneous cloud is reminiscent to an ansatz frequently made in describing Bragg scattering of X-rays on crystals \cite{Coley1995,Weidemueller1998}, where it is found that disorder and impurities do not broaden the angular distribution of the reflected radiation, but increase the background of isotropically distributed diffuse scattering.

8.~The microscopic model can obviously be extended to arbitrary lattice geometries including 3D lattices of any crystalline structure. This is simply due to the fact that before performing a numerical simulation the position of all atoms must be defined, as well as the angle of incidence and the polarization of the probe beam. It makes no difference to the algorithm whether the positions are chosen randomly (to account for disorder) or periodically (to ascribe long-range-order).

\bigskip

\begin{figure}
    \centerline{\scalebox{0.4}{\includegraphics{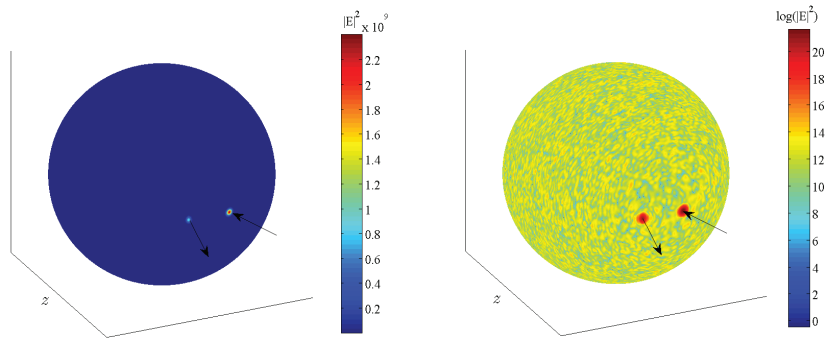}}}
    \caption{Far-field radiation from a one-dimensional optical lattice illuminated with an incident beam
        inclined by $\theta_0=20^\circ$ with respect to the z-axis. The lattice is composed of $100$ disks of
        radius $15\lambda_0$ and thickness $0.04\lambda_0$, spaced by $\lambda_0/2/\cos\theta_0$, with the
        system being composed of $N=8000$ atoms overall. The incident Gaussian beam has waist
        $9\lambda_0$ at the lattice center.}
    \label{fig:Sphere}
\end{figure}

All these effects and experimental constraints are beyond (or have some aspects beyond) the TM formalism, but \emph{all effects} are naturally included in the cooperative scattering model \cite{Courteille2010,Bachelard2011,Bachelard2012}. Effect~7., for example, is confirmed by the appearance of a background of speckle-like scattering well distinct from the solid angle into which the reflected light is scattered, as shown in Fig.~\ref{fig:Sphere}.

\subsection{Suppression of absorption\label{sec:AbsorptionSuppression}}

\begin{figure}
    \centerline{\scalebox{0.6}{\includegraphics{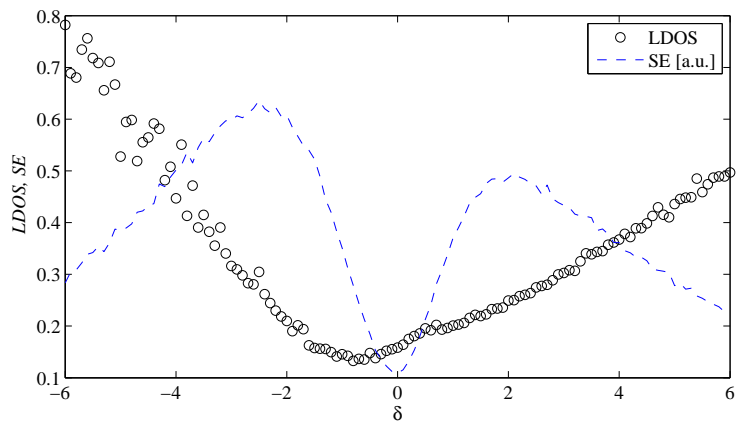}}}
    \caption{Local density of state (circles) and spontaneous emission (dashed line) for a lattice consisting of $10000$
        atoms spread over $200$ disks spaced by $\lambda_0/2$, having thickness $0.04\lambda_0$ and radius
        $15\lambda_0$. The spontaneous emission of the lattice is normalized by that of a single-atom.}
    \label{fig:SE}
\end{figure}

The hallmark of a photonic band is the suppression of spontaneous emission. Models describing the propagation of light inside photonic crystals assign this suppression to a reduction of the density of optical modes available for spontaneous decay \cite{John1983}, as shown in Fig.~\ref{fig:SE}. The suppression of spontaneous emission in a one-dimensional lattice may seem surprising since scattered photons always have the option to escape sideways. However, the reduction of absorption inside a $1D$ band gap can be understood in classical terms \cite{Slama2006}: the standing wave formed by the incident probe beam and the Bragg-reflected light adjusts its phase so that its \textit{intensity nodes coincide with the atomic layers}. In that way absorption is minimized. If the length of the lattice is finite, the contrast of the standing wave is smaller than 1, meaning that the probe light intensity at the locations of the atomic layers does not vanish. Hence, a finite absorption persists even for a perfect but finite lattice with fulfilled Bragg condition, and the photonic band gap cannot be completely dark.

\subsection{Modelling finite-size effects\label{sec:FiniteSizeEffects}}

\begin{figure}[!ht]
    \centerline{\psfig{file=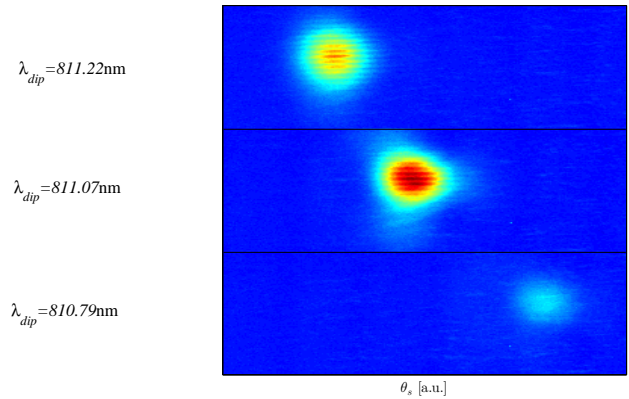,width=10cm}}
    \caption{Variation of the reflection angle with the lattice period when the probe laser is at resonance
        ($\Delta_0=0$). For the chosen angle of incidence, $\theta_0=15.9^\circ$, the Bragg condition is
        fulfilled for the lattice laser wavelength of $\lambda_{lat}=811.07~$nm. The data is recorded using
        the experimental set-up in Ref.~\cite{Slama2005a}.\label{fig:ExpDeviation}}
\end{figure}

To illustrate the superiority of the microscopic model under experimentally realistic conditions we now study the deflection of light reflection in one-dimensional lattices. As observed in~\cite{Slama2005a}, due to the finite lattice size, any deviation of the lattice constant from the Bragg condition for the lattice period leads to an extra inclination of the scattered beam. Fig.~\ref{fig:ExpDeviation} shows the position where the reflected beam impinges on a CCD camera for various lattice constants chosen to satisfy or not the Bragg condition for a given angle of incidence.

\begin{figure}[!ht]
    \centerline{\psfig{file=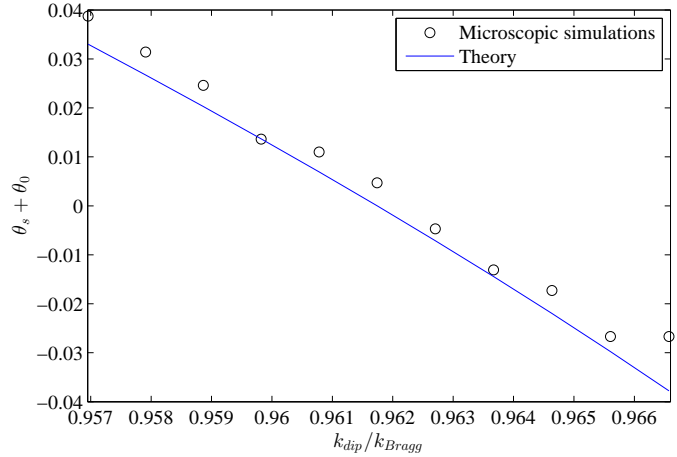,width=10cm}}
    \caption{Deviation of the scattering angle $\theta_s$ from the infinite-slab-Bragg reflection angle
        $-\theta_0$. The blue solid line represents a theoretical expectation~\eqref{eq:line}, the black
        circles are obtained from simulations of the microscopic model \eqref{eqbetajstat}. The
        simulations are realized for a lattice of $8000$ atoms spread over $400$ small disks of thickness
        $0.04\lambda_0$, radius $0.001\lambda_0$ and spaced by $\lambda_{lat}$. The incident laser is
        inclinated by $\theta_0=15.9^\circ$.\label{fig:line}}
\end{figure}

According to~\cite{Slama2005a}, the reflection angle is $\theta_s=\theta_0$ in the case of atomic layers extended to infinity, but should be given by
\begin{equation}
    \cos\theta_0+\cos\theta_s = 2\frac{\lambda_0}{\lambda_{lat}}~\label{eq:line}
\end{equation}
in the case of a {\it one-dimensional line} of point-like scatterers. This prediction (blue solid line in Fig.~\ref{fig:line}) is confirmed by simulations based on our microscopic model~\eqref{eqbetajstat}.

In the intermediate regime, where the slabs are pancake-shaped but have finite radial extension, any deviation of the lattice wavelength from the infinite-slab-Bragg condition $\cos\theta_0=\cos\theta_s=\frac{\lambda_0}{\lambda_{lat}}$ leads to a deviation of the scattered beam, $\theta_s\ne\theta_0$ depending on the aspect ratio of the pancakes. In this regime the Bragg condition~\eqref{eq:line} has to be generalized, as shown in Ref.~\cite{Slama2005a}. On the other hand, as shown above, this physics is contained in our microscopic simulations based on ~\eqref{eqbetajstat}.

\bigskip

\begin{figure}[!ht]
    \centerline{\psfig{file=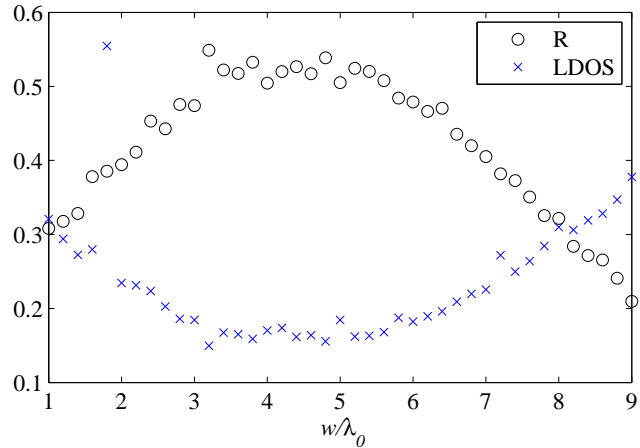,width=10cm}}
    \caption{(color online) Reflection coefficient (dark circles) and LDOS (blue crosses) derived from the
        scalar model as the laser waist is tuned. The simulations are realized for $N=6000$ atoms distributed
        over $N_d=60$ disks of radius $R_0=9\lambda_0$ and thickness $a=0.05\lambda_0$. The laser is incident
        under the angle $\theta_0=2^\circ$ and is resonant, i.e., $\Delta_0=0$. Picture taken from~\cite{Samoylova2013}. \label{fig:waist}}
\end{figure}

Another example demonstrates the utility of the microscopic model. While the TM formalism describes the scattering of a plane-wave from a lattice of infinite transverse size, in reality the probe laser has a Gaussian intensity profile and a finite divergence. In order to study the influence of a finite sized probe laser beam on the PBG, we perform numerical simulations for disks of radius $R_d=9\lambda_0$. Fig.~\ref{fig:waist} shows that the reflectivity reaches a maximum value and the LDOS is minimal for a laser waist in the range of $3-6\lambda_0$. The presence of this maximum is rather intuitive, as if the waist is larger than radial size of the lattice, the laser beam will not be fully intercepted by the lattice, while if it is too small, its divergence is so high that the beam encounters a lower number of disks. In both cases, the contrast of the PBG is reduced.

\bigskip

As a final remark, we note that finite-size effects of the optical lattice have been investigated for a 3D diamond configuration~\cite{Antezza2013}. In this work, the three-dimensional band gap, and more specifically the LDOS, is shown to be strongly affected by either vacancies in the lattice sites or by the finite size.

\section{Conclusions and Perspectives}

The opening of a photonic band gap in an optical lattice is just another manifestation of cooperative scattering. The entire physics of photonics bands is therefore contained within a microscopic collective scattering model that was first described by R.~Dicke \cite{Dicke1954} and earned increasing attention in recent years \cite{Friedberg1973,Scully2006,Courteille2010}. In comparison to the single-photon superradiance, Bragg and Mie scattering, photonic bands occur in an optically dense regime where photons are scattered by different atoms multiple times. Moreover, in contrast to the most frequently studied situation of disordered atomic clouds, photonic bands are a consequence of the order imposed onto the scatterers by the presence of a lattice.

In this paper we aimed at supporting the above assertion by applying the microscopic cooperative scattering model to the only physical system that has been studied in experiment: a one-dimensional optical lattice, producing a stack of pancake-shaped atomic clouds. We proceeded with two approaches. First, the classical TM formalism approach, used in the past to treat 1D optical lattices, has been derived from our cooperative scattering model under certain idealized assumptions. This was followed by numerical simulations of the cooperative scattering model which are compared to theoretical predictions of the TM formalism. However, the cooperative scattering model allows working beyond the TM formalism and the incorporation of a variety of experimental constraints, some of which can have a huge impact on  observations. To grasp the full range of phenomena expected in light scattering from optically dense lattices, a vectorial treatment of the light is necessary. Nevertheless, many features can be determined from a simplified scalar model, which turns out to be an excellent approximation in the case of 1D lattices.

The localization of light and the suppression of spontaneous emission, which represent the holy grail of photonic band gaps, can only be expected for omnidirectional band gaps, which in turn are only supported in three-dimensional lattices. In fact, only few lattice geometries are expected to sustain complete omnidirectional band gaps, for example, diamond-shaped lattices~\cite{Antezza2009} and cubic lattices of driven three-level atoms~\cite{DeshuiYu2011}. Such a lattice would reflect light incident from any direction. Additionally, an excited atom inside the lattice (one could imagine, for example, that the excited state of the atom was reached via spontaneous emission from an upper state) can not emit a photon because the lattice does not offer electromagnetic modes the photon could use. Hence, spontaneous emission is inhibited and the light is trapped in a localized spot. In this respect, there is an interesting analogy to Anderson localization except that in the case of PBGs it is not disorder, but order causing localization.

It is important to stress that the microscopic approach not only is perfectly suited in describing {\it any type of 3D optical lattices}, but it naturally offers the possibility to {\it incorporate disorder and finite-size effects}. It will be the task of upcoming studies to identify the most suitable lattice geometries and other parameters for the detection of omnidirectional band gaps under experimentally realistic conditions. Only then will it be possible to evaluate the potential technological power of optical lattices in molding the flow of light and to compare it with alternative techniques. Indeed, photonic crystals made from dielectric materials are very advanced and powerful devices \cite{Ishizaki2013}. However, these are technically limited as photonic crystals have issues maintaining their periodic order over long distances due to the fabrication process. Here optical lattices are superior because the periodicity is imposed by the lattice lasers and is inherently perfect. Another decisive advantage is the dynamic control possible over optical lattices, which may be ramped up and down or even reshaped in real time.

\section*{Acknowledgements}

This work has been supported by the Funda\c{c}\~{a}o de Amparo \`{a} Pesquisa do Estado de S\~{a}o Paulo (FAPESP) and the Research Executive Agency (Program COSCALI, Grant No. PIRSES-GA-2010-268717).

\begin{appendix}

\section{Appendix: Derivation of the Transfer Matrix $T$\label{Appendix:xn-yn}}

Setting $z=z_n$ and $z=z_n+a$ in (\ref{eq:MultiSlices}) we
obtain, respectively
\begin{eqnarray}\label{bint1}
    \frac{\Omega_0}{\Gamma}e^{ik_{0z}z_n}&=&(2\delta+i)\overline\beta_n(z_n)
    +i\alpha e^{-ik_{0z}z_n}\sum_{m\ge n}\int_{z_m}^{z_m+a} dz'
    e^{ik_{0z}z'}\overline\beta_m(z')\nonumber\\
    &+&i\alpha e^{ik_{0z}z_n}\sum_{m<n}\int_{z_m}^{z_m+a} dz'
    e^{-ik_{0z}z'}\overline\beta_m(z'),
\end{eqnarray}
and
\begin{eqnarray}\label{bint2}
    \dfrac{\Omega_0}{\Gamma}e^{ik_{0z}(z_n+a)}&=&(2\delta+i)\overline\beta_n(z_n+a) \nonumber \\
    &+&i\alpha e^{ik_{0z}(z_n+a)}\sum_{m\le n}\int_{z_m}^{z_m+a} dz'
    e^{-ik_{0z}z'}\overline\beta_m(z')\nonumber\\
    &+&i\alpha e^{-ik_{0z}(z_n+a)}\sum_{m>n}\int_{z_m}^{z_m+a} dz'
    e^{ik_{0z}z'}\overline\beta_m(z'),
\end{eqnarray}
where $\alpha=2\pi\rho_0/(k_0k_{0z})$. Using (\ref{eq:b(z)}),
the integrals give:
\begin{eqnarray}
i\alpha\int_{z_m}^{z_m+a} dz' e^{ik_{0z}z'}\overline\beta_m(z')&=&\left[A x_m +B y_m\right]e^{i k_{0z}z_m},\nonumber\\
i\alpha\int_{z_m}^{z_m+a} dz' e^{-ik_{0z}z'}\overline\beta_m(z')&=&\left[B
e^{i(k_z-k_{0z})a}x_m+A e^{i(k_{z}+k_{0z})a} y_m\right]e^{-i
k_{0z}z_m},\nonumber
\end{eqnarray}
where
\begin{eqnarray}
A&=&\frac{k_z-k_{0z}}{2k_{0z}}\left[1-e^{i(k_{z}+k_{0z})a}\right]\label{A},\\
B&=&\frac{k_z+k_{0z}}{2k_{0z}}\left[1-e^{-i(k_{z}-k_{0z})a}\right],\label{B}
\end{eqnarray}
and (\ref{eq:m0}) and $k_0^2(m_0^2-1)=k_z^2-k_{0z}^2$ were used.
By inserting the above equations into Eqs.(\ref{bint1}) and (\ref{bint2}), we
obtain:
\begin{eqnarray}\label{bint5}
    \frac{\Omega_0}{\Gamma}&=&
    \left[(1+A)x_n+(1+B)y_n\right]e^{-ik_{0z}z_n} \nonumber \\
    &+&\sum_{m<n}\left (Be^{ik_z a}x_m +Ae^{-ik_z a}y_m\right)e^{-ik_{0z}(z_m+a)}\nonumber\\
    &+&e^{-2ik_{0z}z_n}\sum_{m>n}\left(Ax_m+By_m\right)e^{ik_{0z}z_m},
\end{eqnarray}
\begin{eqnarray}\label{bint6}
    \frac{\Omega_0}{\Gamma}&=&
    \left[
    (1+B)e^{ik_z a}x_n+(1+A)e^{-ik_za}y_n\right]e^{-ik_{0z}(z_n+a)} \nonumber \\
    &+&\sum_{m<n}\left (Be^{ik_{z} a}x_m+Ae^{-ik_z a}y_m\right)e^{-ik_{0z}(z_m+a)}\nonumber\\
    &+&e^{-2ik_{0z}(z_n+a)}\sum_{m>n}\left(A x_m+By_m\right)e^{ik_{0z}z_m}.
\end{eqnarray}
Subtraction (\ref{bint6}) from (\ref{bint5}) gives:
\begin{eqnarray}\label{bint7}
    0&=&e^{ik_{0z}z_n}\left[(1+A)-(1+B)e^{i(k_z-k_{0z})a}\right]x_n \nonumber \\
    &+&e^{ik_{0z}z_n}\left[(1+B)-(1+A)e^{-i(k_z+k_{0z})a}\right]y_n\nonumber\\
    &+&\sum_{m=n+1}^N\left(1-e^{-2ik_{0z}a}\right)e^{ik_{0z}z_m}\left(Ax_m+By_m\right),
\end{eqnarray}
where $n=1,\dots,N-1$.  Multiplying (\ref{bint6}) by
$\exp(2ik_{0z}a)$ and subtracting it from (\ref{bint5}) we
obtain:
\begin{eqnarray}\label{bint8}
    \frac{\Omega_0}{\Gamma}\left(1-e^{2ik_{0z}a}\right)&=&
    e^{-ik_{0z}z_n}\left[(1+A)-(1+B)e^{i(k_z+k_{0z})a}\right]x_n \nonumber \\
    &+&e^{-ik_{0z}z_n}\left[(1+B)-(1+A)e^{-i(k_z-k_{0z})a}\right]y_n\nonumber\\
    &+&\sum_{m=1}^{n-1}\left(1-e^{2ik_{0z}a}\right)e^{-ik_{0z}z_m} \nonumber \\
    &\times &\left[Be^{i(k_z-k_{0z})a}x_m+Ae^{-i(k_z+k_{0z})a}y_m,
    \right]
\end{eqnarray}
where $n=2,\dots,N$. In order to find the solution for $N$ slabs
the complete system provided by Eqs. (\ref{bint7}) and
(\ref{bint8}) for the coefficients $x_n$ and $y_n$ should be
solved. The most convenient way to do that is the iterative
method. We introduce
\begin{eqnarray}
K_n&=&e^{ik_{0z}z_n}
\left\{\left[(1+A)-(1+B)e^{i(k_z-k_{0z})a}\right]x_n \right. \nonumber \\
&+&\left. \left[(1+B)-(1+A)e^{-i(k_z+k_{0z})a}\right]y_n\right\}
\end{eqnarray}
and
\begin{equation}
L_n=\left(1-e^{-2ik_{0z}a}\right)e^{ik_{0z}z_n}\left(Ax_n+By_n\right),
\end{equation}
so that (\ref{bint7}) is equivalent to
\begin{equation}
K_n+\sum_{m=n+1}^{N}L_m=0,  \quad n=1,2,...,N-1
\end{equation}
or
\begin{equation}
\label{recurrenceKL} K_{n+1}-L_{n+1}=K_n, \quad n=1,2,...,N-1.
\end{equation}
Analogously, (\ref{bint8}) can be written in the following
form:
\begin{equation}
M_n+\sum_{m=1}^{n-1}N_m=\dfrac{\Omega_0}{\Gamma}\left(1-e^{2ik_{0z}a}\right),
\quad n=2,...,N,
\end{equation}
or
\begin{equation}
\label{recurrenceMN} M_{n+1}=M_n-N_n, \quad n=1,2,...,N-1,
\end{equation}
where
\begin{eqnarray}
M_n&=&e^{-ik_{0z}z_n}\left\{\left[(1+A)-(1+B)e^{i(k_z+k_{0z})a}\right]x_n \right. \nonumber \\
&+&\left. \left[(1+B)-(1+A)e^{-i(k_z-k_{0z})a}\right]y_n\right\}
\end{eqnarray}
and
\begin{equation}
N_n=\left(1-e^{2ik_{0z}a}\right)e^{-ik_{0z}z_n}\left\{
Be^{i(k_z-k_{0z})a}x_n+Ae^{-i(k_z+k_{0z})a}y_n\right\}.
\end{equation}

From Eqs.(\ref{recurrenceKL}) and (\ref{recurrenceMN}) it is
possible to obtain recurrence equations connecting the
coefficients $(x_{n+1},y_{n+1})$ with $(x_{n},y_n)$ in a similar
fashion described in \cite{Boedecker2003}. Using Eqs.(\ref{A}) and
(\ref{B}), a straightforward calculation allows to write
Eqs.(\ref{recurrenceKL}) and (\ref{recurrenceMN}) in the
matrix form:
\begin{equation}\label{SS}
R
\left[%
\begin{array}{c}
  x_{n+1} \\
  y_{n+1} \\
\end{array}%
\right]= S
\left[%
\begin{array}{c}
  x_{n} \\
  y_{n} \\
\end{array}%
\right]
\end{equation}
where
\begin{equation}
R=\left[%
\begin{array}{cc}
  k_{0z}+k_z & k_{0z}-k_z\\
  k_{0z}-k_z & k_{0z}+k_z
   \\
\end{array}%
\right]
\end{equation}
and
\begin{equation}
S=\left[%
\begin{array}{cc}
  (k_{0z}+k_z)e^{i(k_{0z}d+k_z a)} & (k_{0z}-k_z)e^{i(k_{0z}d-k_z a)}\\
  (k_{0z}-k_z)e^{-i(k_{0z}d-k_z a)} & (k_{0z}+k_z)e^{-i(k_{0z}d+k_z a)}
   \\
\end{array}%
\right].
\end{equation}
Multiplying both sides of (\ref{SS}) by the inverse of the left
matrix, we obtain:
\begin{equation}\label{SS2}
\left[%
\begin{array}{c}
  x_{n+1} \\
  y_{n+1} \\
\end{array}%
\right]=T
\left[%
\begin{array}{c}
  x_{n} \\
  y_{n} \\
\end{array}%
\right]
\end{equation}
where
\begin{equation}
\label{Tapp}
T=\left[%
\begin{array}{cc}
   T_{11} & T_{12}\\
   T_{21} & T_{22}
   \\
\end{array}%
\right]
\end{equation}
is the transfer matrix with the elements
\begin{eqnarray}
 T_{11,22}&=&\left[\cos(k_{0z}d)\pm i\frac{k_{0z}^{2}+k_{z}^2}{2k_{0z}k_z}\sin(k_{0z}d)\right] e^{\pm ik_z a},\nonumber \\
 T_{12,21}&=&\pm i\frac{k_{0z}^{2}-k_{z}^2}{2k_{0z}k_z}\sin(k_{0z}d)e^{\mp ik_z a}.\label{eq:coeffT:app}
\end{eqnarray}

We notice that $M_1=\dfrac{\Omega_0}{\Gamma}\left( 1-e^{2ik_{0z}a}\right)$ and $K_N=0$. In addition to these, the iteration equation (\ref{Tn}) at $n=N-1$ provides a link between the coefficients $(x_N,y_N)$ for the $N$th slab with those written for the first slice $(x_1,y_1)$:
\begin{eqnarray}
x_N&=&T_{11}^{N-1}x_1+T_{12}^{N-1}y_1,\nonumber \\
y_N&=&T_{21}^{N-1}x_1+T_{22}^{N-1}y_1,
\end{eqnarray}
so that the coefficients $(x_1,y_1)$ can be derived explicitly. The long but straightforward calculations yield:
\begin{eqnarray}\label{x1}
    x_1&=&\frac{2k_{0z}}{(k_{0z}+k_z)^2-{(k_{0z}-k_z})^2e^{2ik_za}}
    \\ &&\times\frac{[k_{0z}+k_z-({k_{0z}-k_z})r e^{2ik_{0z}d}]\sin(N-1)\phi-t(k_{0z}+k_z)\sin(N-2)\phi}
    {(1-r^2e^{2ik_{0z}d})\sin(N-1)\phi-t\sin(N-2)\phi},\nonumber\\
    y_1&=&\frac{2k_{0z}}{(k_{0z}-k_z)^2-{(k_{0z}+k_z})^2e^{-2ik_za}}
    \\ &&\times\frac{[k_{0z}-k_z-({k_{0z}+k_z})r e^{2ik_{0z}d}]\sin(N-1)\phi-t(k_{0z}-k_z)\sin(N-2)\phi}
    {(1-r^2e^{2ik_{0z}d})\sin(N-1)\phi-t\sin(N-2)\phi},\nonumber\label{y1}
\end{eqnarray}
where $r$ and $t$ are the reflection and transmission coefficients of a single slab given by Eqs.(\ref{Eq:r}) and (\ref{Eq:t}), respectively.

\section{Appendix: Vectorial model\label{Appendix:vectorial}}

In this appendix we summarize the derivation of equations
(\ref{betajk}) for the vectorial scattering model
\cite{Manassah2012}. We consider $N$ two-level atoms, where the
ground and excited states are connected by an electric dipole
transition. We assume that for the $j$th atom the ground state
$|g_j\rangle$ is a singlet and the excited state
$|e_j^{(\alpha)}\rangle$ is a degenerate triplet, with
$\alpha=x,y,z$. For the $j$th atom we define the lowering
operators $\hat\sigma_j^{(\alpha)}=|g_j\rangle\langle
e_j^{(\alpha)}|$ and the electric dipole operator $\mathbf{\hat
d}_j$ with components $\hat
d_j^{(\alpha)}=d\hat\sigma_j^{(\alpha)}\exp(-i\omega_a t)+c.c.$,
where $d$ is the matrix element equal for the three transitions.
The interaction Hamiltonian includes three contributions:
\begin{equation}
H_I=H_L+H_{dd}+H_{rad} \label{Htot},
\end{equation}
where
\begin{equation}
H_L= \frac{d}{2}
\sum_{\alpha}E_{0\alpha}\sum_{j=1}^N\left[\hat\sigma_j^{(\alpha)}e^{i(\Delta_0
t-\mathbf{k}_0\cdot \mathbf{r}_j)}+\mathrm{h.c.}\right]\label{HL}
\end{equation}
is the interaction between the atoms and the incident classical
field with frequency $\omega_0=ck_0$ and wave vector
$\mathbf{k}_0$, where $E_{0\alpha}$ are the electric field
components and $\Delta_0=\omega_0-\omega_a$,
\begin{equation}
H_{dd}= \frac{1}{4\pi\epsilon_0}\sum_{i<j}\frac{1}{r_{ij}^3}
\left[\mathbf{\hat d}_i\cdot \mathbf{\hat d}_j-3(\mathbf{\hat
d}_i\cdot \mathbf{\hat r}_{ij})(\mathbf{\hat d}_j\cdot
\mathbf{\hat r}_{ij}) \right]\label{Hdd}
\end{equation}
is the instantaneous Coulomb interaction between pairs of atoms,
where $\mathbf{r}_{ij}=\mathbf{r}_i-\mathbf{r}_j=
r_{jm}\mathbf{\hat r}_{jm}$, and
\begin{equation}
H_{rad}=\hbar\sum_{j=1}^N\sum_{\alpha}\sum_{\mathbf{k},\mathbf{\hat\epsilon}}g_{k}\hat\epsilon_\alpha
\left(\hat\sigma_j^{(\alpha)}e^{-i\omega_a
t}+\mathrm{h.c.}\right)\left[\hat
a_{\mathbf{k},\mathbf{\hat\epsilon}}\,e^{-i(\omega_k
t-\mathbf{k}\cdot \mathbf{r}_j)}+\mathrm{h.c.}\right]
 \label{Hrad}
\end{equation}
is the interaction between atoms and the vacuum radiation field,
$g_{k}=d(\omega_a^2/2\hbar\epsilon_0 V_{ph}\omega_k)^{1/2}$ is the
single-photon Rabi frequency, $V_{ph}$ is the photon volume and
$\mathbf{\hat\epsilon}$ is the polarization unit vector.
 The dipole-dipole interaction (\ref{Hdd}) can be
written in a more compact form, neglecting the rapidly time
varying terms, as
\begin{equation}
H_{dd}= \frac{1}{8\pi\epsilon_0}\sum_{j\neq
m}\sum_{\alpha,\beta}\left[\hat \sigma_j^{(\alpha)}\hat
\sigma_m^{\dagger(\beta)}+\hat \sigma_j^{\dagger(\alpha)}\hat
\sigma_m^{(\beta)}\right]W_{\alpha,\beta}(\mathbf{r}_{jm}),
\label{Hdd2}
\end{equation}
where
\begin{equation}\label{Wjm}
W_{\alpha,\beta}(\mathbf{r})=\frac{d^2}{r^3}(\delta_{\alpha,\beta}-3\hat
r_{\alpha}\hat r_{\beta}).
\end{equation}
For a weak incident beam we restrict our analysis to the case
where only one photon is absorbed by the atoms. We also consider
the non-rotating-wave terms in the Hamiltonian (\ref{Hrad}), so the
system atoms+photons has the form:
\begin{eqnarray}\label{state:vec}
    |\Psi\rangle&=&\alpha(t)|g_1\dots g_N\rangle |0\rangle_{\mathbf{k}}
    +e^{-i\Delta_0 t}\sum_{j=1}^N\sum_\alpha
    \beta_j^{(\alpha)}(t)|g_1\ldots e_{j}^{(\alpha)},\ldots g_N\rangle|0\rangle_{\mathbf{k}}\nonumber\\
    &+& \sum_{\mathbf{k},\mathbf{\hat\epsilon}}\gamma_{\mathbf{k},\mathbf{\hat\epsilon}}(t)|g_1\dots g_N\rangle
    |1\rangle_{\mathbf{k},\mathbf{\hat\epsilon}}\nonumber\\
    &+&\sum_{\mathbf{k},\mathbf{\hat\epsilon}}\sum_{m\neq n}\sum_{\alpha,\beta}
    \epsilon_{m,n,\mathbf{k},\mathbf{\hat\epsilon}}^{(\alpha,\beta)}(t)|g_1\ldots e_m^{(\alpha)}\ldots e_n^{(\beta)}\ldots
    g_N\rangle|1\rangle_{\mathbf{k},\mathbf{\hat\epsilon}}.
\end{eqnarray}
The first term in ~\eqref{state:vec} corresponds to the initial
ground state without photons, the second term is the sum over the
states where a single atom has been excited by the classical
field. The third term corresponds to the atoms that returned to
the ground state having emitted a photon in the mode $\mathbf{k}$
with polarization $\mathbf{\hat\epsilon}$, whereas the last term
characterizes the presence of two excited atoms and one virtual
photon with `negative' energy. It is due to the counter-rotating
terms in the Hamiltonian (\ref{Hrad}) and disappears when the
rotating wave approximation is made.

With the ansatz (\ref{state:vec}) the Schr\"{o}dinger's equation
reduces to the following system of coupled equations:
\begin{eqnarray}
  \dot\alpha &=& -i\frac{d}{2\hbar}\sum_{j,\alpha}E_{0\alpha}\beta_j^{(\alpha)}e^{-i\mathbf{k}_0\cdot \mathbf{r}_j},
  \label{alpha}\\
  \dot\beta_j^{(\alpha)} &=& i\Delta_0\beta_j^{(\alpha)}- \frac{i}{\hbar}\sum_{m,\beta}W_{\alpha,\beta}(\mathbf{r}_{jm})\beta_m^{(\beta)}
  -i\frac{dE_{0\alpha}}{2\hbar}e^{i\mathbf{k}_0\cdot
  \mathbf{r}_j}\alpha\nonumber\\
  &-&i\sum_{\mathbf{k},\mathbf{\hat\epsilon}}g_{k}\hat\epsilon_{\alpha}e^{-i(\omega_k-\omega_0)t+i\mathbf{k}\cdot \mathbf{r}_j}
\gamma_{\mathbf{k},\mathbf{\hat\epsilon}}\nonumber\\
&-&i\sum_{\mathbf{k},\mathbf{\hat\epsilon}}g_{k}\sum_{m\neq
j,\beta} \hat\epsilon_\beta
e^{-i(\omega_a+\omega_k-\Delta_0)t+i\mathbf{k}\cdot
\mathbf{r}_m}\epsilon_{j,m,\mathbf{k},\mathbf{\hat\epsilon}}^{(\alpha,\beta)},
 \label{beta}\\
  \dot\gamma_{\mathbf{k},\mathbf{\hat\epsilon}}&=& -ig_{k}e^{i(\omega_k-\omega_0)t}\sum_j\sum_\alpha \hat\epsilon_\alpha
  e^{-i\mathbf{k}\cdot \mathbf{r}_j}\beta_j^{(\alpha)}, \label{gammak}
  \\
  \dot\epsilon_{j,m,\mathbf{k},\mathbf{\hat\epsilon}}^{(\alpha,\beta)} &=&
   -ig_{k}e^{i(\omega_k+\omega_a-\Delta_0)t}\hat\epsilon_\alpha\left[
  e^{-i\mathbf{k}\cdot \mathbf{r}_j}\beta_m^{(\beta)}+e^{-i\mathbf{k}\cdot \mathbf{r}_m}\beta_j^{(\beta)}\right]. \label{epsilon}
\end{eqnarray}
By integrating Eqs.(\ref{gammak}) and (\ref{epsilon}) with
$\gamma_{\mathbf{k},\mathbf{\hat\epsilon}}(0)=0$,
$\epsilon_{j,m,\mathbf{k},\mathbf{\hat\epsilon}}^{(\alpha,\beta)}(0)=0$
and substituting them into (\ref{beta}), assuming $\alpha\approx
1$ (linear regime), we obtain:
\begin{eqnarray}
  \dot\beta_j^{(\alpha)} &=& i\Delta_0\beta_j^{(\alpha)}- \frac{i}{\hbar}\sum_{m,\beta}W_{\alpha,\beta}(\mathbf{r}_{jm})\beta_m^{(\beta)}
  -i\frac{dE_{0\alpha}}{2\hbar}e^{i\mathbf{k}_0\cdot
  \mathbf{r}_j}\nonumber\\
  &-&\sum_{\mathbf{k},\mathbf{\hat\epsilon}}g_{k}^2\sum_{m,\beta}\hat\epsilon_{\alpha}\hat\epsilon_{\beta}
  e^{i\mathbf{k}\cdot(\mathbf{r}_j-\mathbf{r}_m})\int_0^t dt'
  \beta_m^{(\beta)}(t')e^{-i(\omega_k-\omega_0)(t-t')}
  \nonumber\\
&-&\sum_{\mathbf{k},\mathbf{\hat\epsilon}}g_{k}^2\sum_{\beta}
\hat\epsilon_\alpha\hat\epsilon_\beta \int_0^t
dt'e^{i(\Delta_0-\omega_k-\omega_a)(t-t')}\nonumber\\
&\times & \left[\sum_{m\neq j}
e^{-i\mathbf{k}\cdot(\mathbf{r}_j-\mathbf{r}_m)}\beta_m^{(\beta)}(t')+(N-1)\beta_j^{(\beta)}(t')
\right].
 \label{beta2}
\end{eqnarray}
The interaction with the vacuum field yields diagonal terms with
$m=j$, whose real part corresponds to the single-atom decay term
and imaginary part to the self-energy shift, and off-diagonal
term with $m\neq j$, which is related  to the atom-atom interaction
mediated by the photon. By separating the two contributions, we
write:
\begin{eqnarray}
  \dot\beta_j^{(\alpha)} &=& i\Delta_0\beta_j^{(\alpha)}- \frac{i}{\hbar}\sum_{m,\beta}W_{\alpha,\beta}(\mathbf{r}_{jm})\beta_m^{(\beta)}
  -i\frac{dE_{0\alpha}}{2\hbar}e^{i\mathbf{k}_0\cdot
  \mathbf{r}_j}\nonumber\\
  &-&\sum_{\beta}\int_0^t d\tau \Lambda_{\alpha,\beta}(\tau)\beta_j^{(\beta)}(t-\tau)\nonumber\\
  &-&\sum_{m\neq j}\sum_{\beta}\int_0^t d\tau K_{\alpha,\beta}(\mathbf{r}_{jm},\tau)
  \beta_m^{(\beta)}(t-\tau),
 \label{beta3}
\end{eqnarray}
where
\begin{eqnarray}
\Lambda_{\alpha,\beta}(\tau)=\sum_{\mathbf{k},\mathbf{\hat\epsilon}}g_{k}^2\sum_{\beta}\mathbf{\hat\epsilon}_{\alpha}
\mathbf{\hat\epsilon}_{\beta}
  \left[e^{-i(\omega_k-\omega_0)\tau}+(N-1)e^{i(\Delta_0-\omega_k-\omega_a)\tau}\right]\nonumber\\
 \label{LL}
\end{eqnarray}
is the self-interaction term and
\begin{eqnarray}
  K_{\alpha,\beta}(\mathbf{r}_{jm},\tau)&=& e^{i\Delta_0\tau}
  \sum_{\mathbf{k},\mathbf{\hat\epsilon}}g_{k}^2\hat\epsilon_{\alpha}\hat\epsilon_{\beta}
  e^{-i\omega_k\tau}\left\{
 e^{i\mathbf{k}\cdot\mathbf{r}_{jm}}e^{i\omega_a\tau}+
 e^{-i\omega_a\tau}e^{-i\mathbf{k}\cdot\mathbf{r}_{jm}}\right\}\nonumber\\
 \label{GG}
\end{eqnarray}
is the inter-atom interaction term. The photon polarization factor
in Eqs.(\ref{LL}) and (\ref{GG}) can be written as
\[
\sum_{\mathbf{\hat\epsilon}}\hat\epsilon_{\alpha}\hat\epsilon_{\beta}=\delta_{\alpha,\beta}-\hat
k_\alpha\hat k_\beta
\]
and the summation over $\mathbf{k}$ can be replaced by
integration:
\[
\int d\mathbf{k}\rightarrow \frac{V_{ph}}{8\pi^3}\int dk
k^2\int_0^{2\pi}d\phi\int_0^\pi d\theta\sin\theta.
\]
The self-interaction term (\ref{LL}) varies on a time scale much
faster than the atomic response, and we can set
$\beta_j^{(\beta)}(t-\tau)\approx \beta_j^{(\beta)}(t)$ and
$t\rightarrow\infty$ in the integral with $\Lambda_{\alpha,\beta}$
in (\ref{beta3}). Then, the real part of this integral yields:
\begin{eqnarray}\label{gamma1}
  \int_0^\infty d\tau
  \mathrm{Re}\Lambda_{\alpha,\beta}(\tau)&=&
  \frac{V_{ph}}{8\pi^3}\int_0^\infty dk
    k^2g_{k}^2\sum_{\beta}\int_0^{2\pi}d\phi\int_0^\pi d\theta\sin\theta\left(\delta_{\alpha,\beta}-\hat
k_\alpha\hat k_\beta\right)\nonumber\\
&\times&\pi\left[\delta(\omega_k-\omega_0)+(N-1)\delta(\omega_k+\omega_a)\right],
\end{eqnarray}
where $\omega_k=ck$. The last term, coming from the
counter-rotating wave terms of the Hamiltonian (\ref{Hrad}), does
not contribute since it corresponds to a negative photon energy
$\omega_k=-\omega_a$, so that
\begin{eqnarray}\label{gamma2}
  \int_0^\infty d\tau
  \mathrm{Re}\Lambda_{\alpha,\beta}(\tau)&=&
  \frac{\Gamma}{8\pi}\sum_{\beta}\int_0^{2\pi}d\phi\int_0^\pi
d\theta\sin\theta\left(\delta_{\alpha,\beta}-\hat k_\alpha\hat
k_\beta\right),
\end{eqnarray}
where $\Gamma=d^2 k_0^3/2\pi\epsilon_0\hbar$. The remaining
angular integration gives
\[
\int_0^{2\pi}d\phi\int_0^\pi
d\theta\sin\theta\left(\delta_{\alpha,\beta}-\hat k_\alpha\hat
k_\beta\right)=\frac{8\pi}{3}\delta_{\alpha,\beta}.
\]
Hence, the energy decay rate of the isolated atom is
$(2/3)\Gamma$. Notice that it differs by a factor $2/3$ from the
scalar decay rate $\Gamma$, which does not include polarization
effects.

The imaginary part of the time integral of
$\Lambda_{\alpha,\beta}$ consists of the self-energy shift of the
atom in the excited state plus the self-energy contribution of the
atom in the ground state. Its effect is an adjustment to the
transition frequency $\omega_a$, which we assume to be already
introduced. Indeed, it cannot be directly treated in our
model that treats the atoms as point particles, so it is
disregarded in the present approach. With these results,
(\ref{beta3}) becomes:
\begin{eqnarray}
  \dot\beta_j^{(\alpha)} &=& \left(i\Delta_0-\frac{\Gamma}{3}\right)\beta_j^{(\alpha)}
  +\frac{1}{i\hbar}\sum_{m,\beta}W_{\alpha,\beta}(\mathbf{r}_{jm})\beta_m^{(\beta)}
  -i\frac{dE_{0\alpha}}{2\hbar}e^{i\mathbf{k}_0\cdot
  \mathbf{r}_j}\nonumber\\
  &-& \sum_{m\neq j}\sum_{\beta}\int_0^t d\tau
K_{\alpha,\beta}(\mathbf{r}_{jm},\tau)
  \beta_m^{(\beta)}(t-\tau).
 \label{beta4}
\end{eqnarray}
Now we face the evaluation of the time-dependent kernel
$K_{\alpha,\beta}(\mathbf{r}_{jm},\tau)$ defined by (\ref{GG}).
Summing over polarization and replacing $\mathbf{k}$ by
integration, we obtain:
\begin{eqnarray}
  K_{\alpha,\beta}(\mathbf{r}_{jm},\tau)&=& \frac{V_{ph}}{(2\pi)^3}e^{i\Delta_0\tau}
  \int_0^\infty dk k^2 g_{k}^2 e^{-i\omega_k\tau}\int_0^{2\pi}d\phi\int_0^\pi
d\theta\sin\theta\nonumber\\
  &\times & \left(\delta_{\alpha,\beta}-\hat k_\alpha\hat
k_\beta\right) \left\{
 e^{ikr_{jm}\cos\theta+i\omega_a\tau}+
 e^{-ikr_{jm}\cos\theta-i\omega_a\tau}\right\},\nonumber\\
 \label{KK}
\end{eqnarray}
where $r_{jm}=|\mathbf{r}_{jm}|$. Notice that the polar angle
$\theta$ depends on the orientation of the vector
$\mathbf{r}_{jm}$, but for simplicity of notation we have omitted
the subscript $(j,m)$. The integration over the azimuth angle
$\phi$ yields:
\begin{eqnarray}\label{avephi}
\frac{1}{2\pi}\int_0^{2\pi}d\phi\left(\delta_{\alpha,\beta}-\hat
k_\alpha\hat
k_\beta\right)=\frac{1}{2}(1+\cos^2\theta)\delta_{\alpha,\beta}+
\frac{1}{2}(1-3\cos^2\theta)\hat r_{\alpha}\hat
r_{\beta}.\nonumber\\
\end{eqnarray}
Using (\ref{avephi}) in (\ref{KK}) and integrating over
$\theta$,
\begin{eqnarray}
  K_{\alpha,\beta}(\mathbf{r},\tau)&=& \frac{\Gamma c^2}{\pi\omega_a}e^{i\Delta_0\tau}\cos(\omega_a\tau)
  \int_0^\infty dk k
  e^{-ick\tau}\left\{
  \left(\delta_{\alpha,\beta}-\hat r_\alpha\hat
r_\beta\right)\frac{\sin(kr)}{kr}\right.\nonumber\\
  &+&\left.\left(\delta_{\alpha,\beta}-3\hat r_\alpha\hat
r_\beta\right)\left(\frac{\cos(kr)}{(kr)^2}-\frac{\sin(kr)}{(kr)^3}\right)\right\},
 \label{KKK}
\end{eqnarray}
where we used the definitions of $g_k^2$ and $\Gamma$. Then, by
integrating by parts the last term in the braces of
(\ref{KKK}) and assuming
$\exp(i\Delta_0\tau)\cos(\omega_a\tau)\approx\cos(\omega_0\tau)$
since $\Delta_0\ll\omega_0$ and $\omega_a\sim\omega_0$, we obtain:
\begin{eqnarray}
  K_{\alpha,\beta}(\mathbf{r},\tau)&=& \frac{\Gamma c^2}{\pi\omega_0}
  \cos(\omega_0\tau)\left\{\frac{1}{r}\left(\delta_{\alpha,\beta}-\hat r_\alpha\hat
  r_\beta\right)\int_0^\infty dk
  e^{-ick\tau}\sin(kr)\right.\nonumber\\
  &+&\left.\frac{1}{r^3}\left(\delta_{\alpha,\beta}-3\hat r_\alpha\hat
r_\beta\right)\left(ic\tau\int_0^\infty dk
  e^{-ick\tau}\frac{\sin(kr)}{k}-r\right)\right\}.
 \label{K4}
\end{eqnarray}
Equation (\ref{K4}) contains the following integrals:
\begin{eqnarray}
\int_0^\infty dk
  e^{-ick\tau}\sin(kr)&=&\frac{1}{2}\lim_{\epsilon\rightarrow 0^+}
  \left[\frac{1}{r-c\tau+i\epsilon}+\frac{1}{r+c\tau-i\epsilon}\right],\label{int1}\\
\int_0^\infty dk
  e^{-ick\tau}\frac{\sin(kr)}{k}&=&\frac{1}{2i}
  \left[\ln\left|\frac{\tau+r/c}{\tau-r/c}\right|+i\pi\theta(r-c\tau)\right],\label{int2}
\end{eqnarray}
where $\theta(x)=1$ for $x>0$ and $\theta(x)=0$ for
$x<0$. We observe that $K_{\alpha,\beta}(\mathbf{r},\tau)$ depends
on $\tau\pm r/c$. Assuming that $\beta_j^{(\alpha)}$ varies on a
characteristic time much longer than $r_{jm}/c$ between any two
atoms $j,m$ in the sample, $\beta_j^{(\beta)}(t-\tau)$ can be
replaced by $\beta_j^{(\beta)}(t)$ during the integration time and
the upper integration limit can be extended to infinity. The
integration over $\tau$ of $K_{\alpha,\beta}(\mathbf{r},\tau)$
requires the further evaluation of the integrals:
\begin{eqnarray}\label{int3}
    &&\int_0^\infty d\tau
    \cos(\omega_0\tau)\left[\frac{1}{r-c\tau+i\epsilon}+\frac{1}{r+c\tau-i\epsilon}\right]
    =\int_{-\infty}^\infty d\tau
    \frac{\cos(\omega_0\tau)}{r-c\tau+i\epsilon}=\frac{\pi}{ic}e^{ik_0r}\nonumber\\
\end{eqnarray}
and
\begin{eqnarray}\label{int4}
    &&\int_0^\infty d\tau\cos(\omega_0\tau)
    \left\{\frac{c\tau}{2}
    \left[
    \ln\left|\frac{\tau+r/c}{\tau-r/c}\right|+i\pi\theta(r-c\tau)
    \right]-r\right\}\nonumber\\
    &&=\frac{i\pi}{2ck_0^2}\left[
    (1-ik_0r)e^{ik_0r}-1\right].
\end{eqnarray}
Using Eqs.(\ref{int1})-(\ref{int4}), we obtain:
\begin{eqnarray}\label{GK5}
    K_{\alpha,\beta}(\mathbf{r})&=&
    \frac{\Gamma}{2}
  \left\{\frac{1}{i(k_0r)}\left(\delta_{\alpha,\beta}-\hat r_\alpha\hat
  r_\beta\right)e^{ik_0r}\right.\nonumber\\
  &+&\left.\frac{i}{(k_0r)^3}\left(\delta_{\alpha,\beta}-3\hat r_\alpha\hat
r_\beta\right)\left[
    (1-ik_0r)e^{ik_0r}-1\right]\right\}.
\end{eqnarray}
The contribution from the non-oscillating term in (\ref{GK5})
cancels the electrostatic term $-(i/\hbar)W_{\alpha,\beta}$ in
(\ref{beta4}). Reintroducing the subscript $i,j$, the equation
for $\beta_j^{(\alpha)}$ is finally obtained:
\begin{eqnarray}
  \dot\beta_j^{(\alpha)} &=& \left(i\Delta_0-\frac{\Gamma}{3}\right)\beta_j^{(\alpha)}
  -i\frac{dE_{0\alpha}}{2\hbar}e^{i\mathbf{k}_0\cdot
  \mathbf{r}_j}
  -\frac{\Gamma}{2}\sum_{m\neq j}\sum_{\beta}G_{\alpha,\beta}(\mathbf{r}_{jm})
  \beta_m^{(\beta)},\nonumber\\
 \label{beta5}
\end{eqnarray}
where
\begin{eqnarray}\label{Gfinal}
    G_{\alpha,\beta}(\mathbf{r})&=&
    \frac{e^{ik_0r}}{ik_0r}
  \left\{\left[\delta_{\alpha,\beta}-(\hat r_\alpha\hat r_\beta\right]+\left[\delta_{\alpha,\beta}-3\hat r_\alpha\hat r_\beta\right]
  \left[\frac{i}{k_0r}-\frac{1}{(k_0r)^2}\right]\right\}.\nonumber\\
\end{eqnarray}

\section{Appendix: Vectorial field\label{Appendix:field}}

The expression for the radiation field can be obtained from the
Maxwell equations in the presence of a polarization $\mathbf{P}$:
\begin{eqnarray}
  \nabla\times \mathbf{E} &=& -\frac{\partial \mathbf{B}}{\partial t} \label{eqM1},\\
  \nabla\times \mathbf{B} &=& \mu_0\frac{\partial \mathbf{D}}{\partial t} \label{eqM2},\\
  \nabla\cdot \mathbf{D} &=& 0, \label{eqM3}
\end{eqnarray}
where $\mathbf{D}=\epsilon_0 \mathbf{E}+\mathbf{P}$. From
Eqs.(\ref{eqM1}) and (\ref{eqM2}) we obtain:
\begin{eqnarray}
  \nabla\times\nabla\times \mathbf{E} &=& -\frac{\partial }{\partial t}
  (\nabla\times \mathbf{B})= -\mu_0\frac{\partial^2 }{\partial t^2}
  (\epsilon_0 \mathbf{E}+\mathbf{P}),\label{eqM12}\\
  \nabla\cdot \mathbf{E} &=& -\frac{1}{\epsilon_0}\nabla\cdot \mathbf{P}.\label{eqM4}
\end{eqnarray}
Since $\nabla\times\nabla\times \mathbf{E}=\nabla(\nabla\cdot
\mathbf{E})-\nabla^2 \mathbf{E}$ and $\epsilon_0\mu_0=c^{-2}$,
then
\begin{eqnarray}
  \nabla^2 \mathbf{E} -\frac{1}{c^2}\frac{\partial^2 \mathbf{E}}{\partial t^2}&=&
  \frac{1}{\epsilon_0 c^2}\frac{\partial^2 \mathbf{P}}{\partial t^2}-\frac{1}{\epsilon_0}\nabla(\nabla\cdot \mathbf{P})
  \label{eqM4b}.
\end{eqnarray}
For a monochromatic field with frequency $\omega_0=ck_0$ we have:
\begin{eqnarray}
  \left(\nabla^2 +k_0^2\right)\mathbf{E}&=&-\frac{k_0^2}{\epsilon_0}\left[
  \mathbf{P}+\frac{1}{k_0^2}\nabla(\nabla\cdot \mathbf{P})\right].
  \label{eqM5}
\end{eqnarray}
Using (\ref{green}), the solution of (\ref{eqM5}) is
\begin{equation}\label{Er}
\mathbf{E}(\mathbf{r})=i\frac{k_0^3}{4\pi\epsilon_0}\int
d\mathbf{r}' G(|\mathbf{r}-\mathbf{r}'|)
\left[\mathbf{P}(\mathbf{r}')+\frac{1}{k_0^2}\nabla(\nabla\cdot
\mathbf{P}(\mathbf{r}'))\right]
\end{equation}
or for each component:
\begin{equation}\label{Er21}
E_\alpha(\mathbf{r})=i\frac{k_0^3}{4\pi\epsilon_0}\sum_{\beta}\int
d\mathbf{r}' G(|\mathbf{r}-\mathbf{r}'|)
\left[\delta_{\alpha,\beta}+\frac{1}{k_0^2}\frac{\partial^2}{\partial
x'_\alpha\partial x'_\beta}\right]P_\beta(\mathbf{r}').
\end{equation}
By integrating by parts and using (\ref{GvGs}), we obtain:
\begin{equation}\label{Er2}
E_\alpha(\mathbf{r})=i\frac{k_0^3}{4\pi\epsilon_0}\sum_{\beta}
\int d\mathbf{r}' G_{\alpha,\beta} (\mathbf{r}-\mathbf{r}')
P_{\beta}(\mathbf{r}').
\end{equation}
Taking $P_\alpha(\mathbf{r})=-d\rho\beta^{(\alpha)}(\mathbf{r})$
and returning to a discrete distribution of atoms with position
$\mathbf{r}_j$, the scattered field components at the position
$\mathbf{r}$ are
\begin{equation}\label{Ebeta}
E_\alpha(\mathbf{r})=-i\frac{dk_0^3}{4\pi\epsilon_0}\sum_{\beta}\sum_{m}
G_{\alpha,\beta}(\mathbf{r}-\mathbf{r}_{m}) \beta^{(\beta)}_m.
\end{equation}
Combining Eqs.\eqref{Ebeta} and \eqref{beta5}, we find that the
stationary atomic polarization components are given by:
\begin{equation}\label{betavsE}
   \beta_j^{(\alpha)}=\frac{d}{\hbar\left(\Delta_0+i\Gamma/3\right)}
   \left[\frac{E_{0\alpha}}{2}e^{i\mathbf{k}_0\cdot \mathbf{r}_j}+E_\alpha(\mathbf{r}_j)\right].
\end{equation}

\end{appendix}

\bibliographystyle{ws-book-har}    
\bibliography{BiblioCollectiveScattering}      


\end{document}